\documentclass[reprint,aip,jcp,showkeys,superscriptaddress]{revtex4-2}
 
\usepackage{graphicx,bm,xcolor,microtype,multirow,amscd,amsmath,amssymb,amsfonts,longtable,wrapfig,txfonts,soul,dsfont}
\usepackage{adjustbox}
\usepackage[normalem]{ulem}

\newcommand{\red}[1]{\textcolor{black}{#1}}

\usepackage[utf8]{inputenc}
\usepackage[T1]{fontenc}
\usepackage{txfonts}
\usepackage{siunitx}

\usepackage[
	colorlinks=true,
    citecolor=blue,
    breaklinks=true
	]{hyperref}
\newcommand{\br}{\mathbf{r}}
\newcommand{\dd}[0]{\text{d}}

\begin{document}

\title{ Efficient analytic continuation approach to Bethe-Salpeter excitation spectra in selected energy windows  }

  \author{Ivan Duchemin}
\affiliation{Univ. Grenoble Alpes, CEA, IRIG-MEM-L\_Sim, 38054  Grenoble, France}
\email{ivan.duchemin@cea.fr}
\author{Xavier Blase}
\affiliation{ Univ. Grenoble Alpes, CNRS, Institut N\'{e}el, F-38042 Grenoble, France }

\date{\today}

\begin{abstract}

We explore the merits of building the Bethe-Salpeter absorption spectrum in a specific energy range using analytic continuation techniques. Specifically, we calculate iteratively a few $\bar{\bar \alpha}(z_k)$ polarizability tensors for a coarse set of $(z_k)$ frequencies in the complex-plane. These data allow constructing a continued-fraction representation for $\bar{\bar{\alpha}}(z)$ that is used to calculate the absorption spectrum close to the real energy axis in the desired energy range.   The number and location of these sampling complex frequencies are discussed.
The importance of building a continued-fraction representation of the full polarizability tensor  with matrix-valued coefficients is emphasized. We  show how to extract the poles of the continued fraction as a tool for analyzing the resulting spectra. We study as examples the valence excitations of a paradigmatic dipeptide, the C$_{60}$ fullerene and its PCBM derivative, together with the description of the surface plasmon resonance of the Ag$_{20}$ silver cluster. Further, the high-energy C$_{60}$ X-ray absorption spectrum is explored.   
\end{abstract}


\maketitle

\section{Introduction}

The Bethe-Salpeter equation (BSE) formalism, \cite{Salpeter1951,Csanak1971,Strinati1982,Strinati1988}  with a nonlocal exchange-correlation kernel that derives from the $GW$ self-energy, \cite{Hedin1965} has been shown in solid-state physics and molecular chemistry to stand as a valuable alternative to time-dependent Density Functional Theory (TD-DFT). The  BSE kernel allows in particular   dealing properly with the needed non-locality of the electron-hole interaction in the case of  extended Wannier excitons, \cite{Albrecht1998,Rohlfing1998,Benedict1998,Botti2004,Sottile2007} charge-transfer excitations in molecular crystals \cite{Hummer2004,Cudazzo2012} or donor-acceptor complexes, \cite{Rocca2010,Blase2011,Duchemin2012,Baumeier2012,Hirose2017,Forster2022} Rydberg states, etc. Focusing on  gas phase molecular systems,  large benchmarks against higher-level quantum chemistry calculations, performed with identical geometries and basis sets, allowed assessing the accuracy of the BSE formalism for excitations of different nature. \cite{Jacquemin2015,Bruneval2015,Jacquemin2017,Gui2018,Holzer2018,Nguyen2019,Rangel2017,Liu2020,Loos2021,Knysh2024} 
 In terms of computational cost,  the BSE eigenvalue problem can be expressed in  a matrix representation very similar to the Casida's equation for TD-DFT \cite{Casida1995} in the product-space of occupied-to-unoccupied (electron-hole) Bloch states or molecular orbitals. As such, once the $GW$ calculations performed, adiabatic BSE calculations can be performed using the same techniques as that developed for TD-DFT. 

The standard approach to BSE calculations proceeds by calculating  the  eigenstates of the BSE  Hamiltonian in the electron-hole basis, obtaining  not only the excitation energies but also the associated excitonic eigenstates.
The knowledge of the two-body (electron-hole) eigenstates is an important asset that allows  analyzing the physical nature of a given excitation (Wannier, Frenkel, charge-transfer, Rydberg, collective plasmon mode, etc.)  Further, dark states with null oscillator strength are available even though not detectable in the absorption spectrum. Since the particle-hole product space scales quadratically with the size $N$ of the system, the search for all eigenstates  leads to an $\mathcal{O}(N^6)$ process. As a result, iterative approaches,  such as the Davidson technique,   are preferred to obtain selectively the low-lying eigenstates of the BSE excitation spectrum. These iterative approaches scale  typically as $\mathcal{O}(N_{\text{exc}}N^4)$ where $N_{\text{exc}}$ is the number of desired low-lying excitations. Still, the number $N_{\text{exc}}$ of excitations  in a given energy window \red{ scales quadratically with  system size, \cite{quadratic} } leading to a densification of eigenstates  even in the low energy range.

The iterative search for the low-lying BSE  eigenstates becomes more expensive when high-energy excitations are needed. This is e.g. the case of core-to-valence transitions lying above the very large number of valence excitations. While  BSE calculations of the X-ray absorption spectrum (XAS) including both (semi)core and valence hole states in the product-space  have been performed for semicore levels in elemental solids, \cite{Urquiza2023} or small to medium molecular systems, \cite{Noguchi2015,Yao2022}  an efficient approach consists in   reducing the manifold of occupied (initial) states to the desired core states, namely removing in particular the valence-to-unoccupied blocks from the BSE Hamiltonian. \cite{Olovsson2009,Laskowski2010,Unzog2022,Kehry2023} This popular core-valence separation (CVS) scheme \cite{Cederbaum1980,Herbst2020} leads to bringing the targeted core excitations to the bottom of the excitation spectrum for this tailored BSE sub-matrix. Besides absorption, the X-ray emission spectrum of small molecules was also studied at the $GW$ and BSE level. \cite{Aoki2019} 

Alternatively, techniques aiming at obtaining directly the optical absorption spectrum, without the knowledge of the excitonic eigenstates, have been developed at the TD-DFT \cite{Walker2006,Rocca2008}  and BSE \cite{Vinson2011,Benedict1998,Rocca2010,Gruning2011,Gilmore2015,Ljungberg2015}  levels on the basis of calculating  the needed matrix elements of the two-body Hamiltonian resolvent, that we  loosely write $L(\omega) \propto ( \omega \mathcal{I} - \mathcal{H}^{\text{BSE}} )^{-1}$ where $\mathcal{H}^{\text{BSE}}$ is the BSE Hamiltonian. The notations will be consolidated below. This is achieved using typically Haydock recursion techniques. \cite{Haydock1972,Haydock1975,Haydock1980}
Lanczos iterative schemes were designed to determine iteratively the coefficient of a continued fraction representation of the resolvant matrix elements. From this continued fraction representation, the absorption at any energy can be calculated in principle. In practice, these approaches require more and more iterations, involving Hamiltonian-on-vector multiplications, \red{as one wishes to} increase the energy window described with accuracy, starting from the lowest excitation energies. \cite{Benedict1998,Rocca2010,Gruning2011,Ljungberg2015} \red{As in the case of recursive diagonalization, such an approach cannot tackle selectively a given energy window without converging all excitations located at lowest energy. As a result, obtaining selectively the absorption spectrum associated with high-lying excitations, such as core or semicore excitations, or resonances immersed in a quasi-continuum of states, may prove very demanding.} In the case of core excitations,  the CVS partitioning scheme could be adapted  by restricting the  Lanczos vectors to the product of core states with unoccupied states,  converging selectively the spectrum in the X-ray absorption energy range. \cite{Shirley1998,Vinson2011,Gilmore2015} \red{ }

As a recent alternative, the projection $ {L}(\omega)D$, where $D$  is the vector  collecting the dipole matrix elements in electron-hole transition space,  was calculated directly at specific frequencies as the solution $X(\omega)$ of the  system $\; (\omega \mathcal{I} -\mathcal{H}^{\text{BSE}} )X(\omega) = D$ solved iteratively. \cite{Kehry2020,Kehry2023}  This elegant approach allows calculating the polarizability tensor at a given energy, but does not provide a functional form that can be used in some extended energy range. The BSE core level absorption spectrum of small molecular systems was calculated beyond the CVS approximation, including scalar relativistic and spin-orbit coupling.   The polarizability tensor was calculated for a fine energy grid (typical spacing 0.05~eV) in the desired energy range. The sampling frequencies were attributed a small imaginary part to mimic the experimental broadening, leading to the wording of damped-response formulation.   

In the present study, we merge this latter approach with analytic continuation techniques by calculating $ {L}(z)D$ at specific energies $\lbrace z_k \rbrace$ in the complex plane so as to build a continued fraction representation of the polarizability tensor that can be analytically continued close to the real axis. We explore in particular to which extent a very coarse energy grid in the complex plane allows   capturing by continuation the main features of the absorption spectrum in the targeted energy range. After considering for validation low-energy valence excitations in selected molecular systems, we   explore the  case of core  excitations taking the   $C_{60}$ fullerene as example. We show further that this approach is particularly suited to capture very efficiently the envelope of broad structures composed of a near continuum of excitations, considering the important example of localized surface plasmon resonances (LSPR) in metallic nanoparticles. Rather than constructing independently the analytic continuation through \red{Pad{\'e} approximants } of each matrix elements of the 3$\times$3 polarizability tensor, we show that a single continued fraction representation of the full polarizability tensor, involving matrix-valued coefficients, leads to much more accurate spectra for a given number of reference $\lbrace z_k \rbrace$ sampling frequencies. Further, we show how to extract the poles and associated residues of a continued-fraction as a tool for analyzing the absorption spectra. An efficient preconditioner for the used generalized minimal residual  (GMRES) method  is introduced.

We leave this Introductory Section by mentioning alternative efficient strategies, such as making the trial (Krylov) eigenvectors orthogonal to lower lying eigenstates,  \cite{Liang2011,Hillenbrand2025} or propagating in time an initial wavepacket through a time-dependent BSE formalism, reconstructing the absorption spectrum by Fourier transform. \cite{Schmidt2003,Fuchs2008,Attaccalite2011,Elliott2021,Marek2025} 
In all cases, the application of the BSE Hamiltonian onto a  vector in 
transition space is the central operation. Strategies designed to accelerate this projection  (stochastic techniques,\cite{Rabani2015,Bradbury2022} Wannier localization, \cite{Merkel2024} reduction of  $\mathcal{H}^{\text{BSE}}$   to some (occupied)$\times$(unoccupied) product subspace, etc.), will not be dealt with in the present study.  
Leaving aside model BSE Hamiltonians, such accelerating techniques can be combined with the ideas discussed below.


\section{Analytic continuation of spectral functions} 

We focus in this section on the use of the analytic continuation technique applied to response or spectral functions in general. We first describe Thiele's continued fractions   and discuss their extension to matrix-valued interpolants.  We also describe how to extract spectral representations from these interpolants, namely the associated poles and residues. We briefly discuss construction details such as frequency sampling and complex plane symmetries. The case of the Bethe-Salpeter dynamical polarizability  will be discussed in a second section as a specific    application. 

\subsection{The analytic continuation technique}

Analytic continuation is a  complex-analysis technique intended to extend the domain of definition of a given analytic function. In numerical applications, when the proper functional form of the target function is respected (such as poles structure and asymptotic behavior), analytic continuation often succeeds in providing accurate values of a function in such extended domains. Although there are several techniques to perform analytic continuation (such as Pad\'{e} approximants,~\cite{Lorentzen2010} barycentric interpolation with the AAA algorithm,~\cite{AAA}  or the Krylov-based RKFit approach~\cite{RKFIT}),  we choose here to rely on Thiele's interpolation formula. \cite{Thiele1909}
This formula defines a rational function $f(z)$ from a finite set of inputs $\{z_i\}$ and their function values $\{f(z_i)\}$. Given $2n$ sampling $\{f(z_i)\}$ values, the formula produces a continued fraction that corresponds to a Pad\'{e} approximant of order $[n-1/n]$, namely the ratio of an order-(n-1) polynomial by an order-n polynomial. 
This functional form  ensures that the structure of the targeted response functions (sum of simple poles) is respected. Within the many-body perturbation-theory (MBPT) framework, this approach was first proposed by Vidberg in Ref.~\citenum{Vidberg77} for the continuation of the self-energy from the imaginary to the real energy axis. In its general form, the resulting continued fraction reads:
\begin{equation}
f^{AC}(z) = b_0 + \cfrac{a_0}{b_1+ \cfrac{(z-z_1)a_2}{b_2 + \cfrac{(z-z_2)a_3}{ b_3 + \cdots }}}
\end{equation}
where the $a_i$ and $b_i$ coefficients are interdependent through the  equivalence transformation. For instance, in its original work,~\cite{Vidberg77} Vidberg's
construction scheme sets automatically all $b_{i>0}=1$. Here the superscript ``{A}C" stands for Analytic Continuation to emphasize that $f^{AC}(z)$  will be used to approximate the function $f(z)$ away from the reference $\{z_i,f(z_i)\}$ data points.

When constructed carefully, such approximants have been shown to provide very accurate results (see, for example, Refs.~\citenum{Tan_2001,Celis_2024}). They also have the advantage of being adaptable to accommodate the eventual symmetries of the target function, as further demonstrated. 
It is also worth anticipating here that Thiele interpolants are built iteratively, and that resulting interpolants may depend on the order of inclusion of the sampling points. The computation of inverse differences and successive convergents involved in the construction of the Thiele continued fraction may also suffer from numerical instabilities,~\cite{CUYT1988} with in the worst case an exponential loss of precision occurring during the evaluation of the inverse differences.~\cite{MORRIS1980}  However, a careful selection of the inclusion order of the sampling generally circumvents such instabilities.~\cite{Celis_2024} \red{ A similar strategy was enforced for $GW$ calculations exploiting analytic continuations techniques. \cite{Duchemin2020,Barrueta2023} } These aspects will be discussed below.

\subsection{Continued fraction for scalar response function}  

In practice, in the case of a dynamical response function which should tends to 0 as $z\to\pm\infty$, a first reasonable choice is to set $b_0=0$ and restrict the continued fraction to an even number of reference function values $N_f=2n$, recovering the correct (1/z) asymptotic behavior of the function. Also, for reasons that will become clear in the following section and contrary to Vidberg's initial choice, we will consider the case where all $a_i=1$. Using standard  notations, the resulting continued fraction reads:
\[
f^{AC}(z) = \cfrac{1}{b_1+} \; \cfrac{(z-z_1)}{b_2 +} \; \cfrac{(z-z_2)}{ b_3 +}  \cdots \cfrac{(z-z_{2n-1})}{ b_{2n} }
\]
Each $(z-z_i)$ factor acts as a stop that cuts the tail of the fraction for  $z=z_i$, allowing to define recursively the $b_i$ coefficients so that $f^{AC}(z_i)$  matches exactly the reference $f(z_i)$ value. The recursion procedure can be made more explicit by introducing the partial denominators $g_i(z)$ as:
\[
g_i(z) = \cfrac{1}{b_i+(z-z_i)g_{i+1}(z)}\,,\quad f^{AC}(z) = g_1(z)
\]
The recursion starts by setting the initial partial denominators to the calculated reference values $g_1(z_j)=f(z_j)$ for all samples $j\in(1,\dots,2n)$ and deduce the term $b_1$ to match the resulting continuant in $z=z_1$:
\[
 f^{AC}(z_1) = f(z_1) \rightarrow b_1 =  g_1(z_1)^{-1}   
\]
From the knowledge of $b_1$, one can propagate the reference values $g_1(z_j)$ to the next partial denominators $g_2(z_j)$: 
\[
 g_{2}(z_{j>1}) = \cfrac{g_1(z_j)^{-1}-g_1(z_1)^{-1}}{z_j-z_1}  
\]
Matching the continuant in $z=z_2$ brings then
\[
 f^{AC}(z_2) = f(z_2) \rightarrow  b_2 = g_{2}(z_{2})^{-1} 
\]
The process is repeated iteratively in order to include all the sampling points. At stage $i$ of the recursion, we  have likewise :
\begin{align}
b_i = g_i(z_i)^{-1}\;, \quad g_{i+1}(z_{j>i}) = \cfrac{g_i(z_j)^{-1}-g_i(z_i)^{-1}}{z_j-z_i} 
\label{eqn:thiele4}
\end{align}
where the condition on $b_i$ enforces that   $f^{AC}(z_i)=f(z_i)$, while the $g_{i+1}(z_j)$ are set so as to propagate the necessary conditions on $f(z_j)$ for the remaining reference points $z_{j>i}$.

\subsection{Full tensor analytic continuation} 

The response functions we target generally come from the resolvent of an Hamiltonian operator, noted $\mathcal{H}$ here. In many cases one has to deal with a matrix representation of the response operator, expressed in a given subspace of interest.
More explicitly, we may consider a generic matrix representation ${\bf f}$ of the resolvent, with scalar entries $f_{pp'}(z) = \langle p |\, \big(\mathcal{H} -z\mathds{I}\big)^{-1} | p' \rangle$ limited to some reduced basis $\big\{\,|p\rangle\,\big\}$.  While the analytic continuation of individual $f_{pp'}$ matrix elements of the resolvent can be performed, it is advantageous to notice here that all matrix entries share a common pole structure, namely the eigenvalues of $\mathcal{H}$. In these cases, a simultaneous continuation of the whole response tensor may then prove superior:
i) working on the full tensor at once imposes the constraint that the analytic continuation will yield poles at the exact same location for all individual matrix elements, contrary to an independent analytic continuation for each components which does not warrant that the poles from different entries will be consistent.
ii) as a connected argument, the continuation of a $p\times p$ tensor through $2n$ estimations will allows reproducing a total of $n\times p$ poles, leading to a finer structure at the matrix entry level than what is permitted by the $n$ poles resulting from the scalar approach.

Formally, we want to construct the continued fraction using now matrix-type coefficients ${\bf b}_i$ instead of scalars: 
\begin{align}
 {\bf f}^{AC}(z) = \cfrac{1}{   {\bf b}_1 + } \; \cfrac{(z-z_1)}{  {\bf b}_2 + } \; \cfrac{(z-z_2)}{ {\bf b}_3 +}  \cdots \cfrac{(z-z_{2n-1})}{  {\bf b}_{2n} }
\end{align}
where we put in bold matrix-valued coefficients or operators expressed in a basis. The reason for choosing the convention $a_i=1$ (or equivalently ${\bf a}_i = \mathds{1}$) becomes clear at this stage: it allows preserving the tensor symmetries of the resolvent (in the case of a real-symmetric Hamiltonian, it should be anti-hermitian). Contrary to the $b_{i>0}=1$ case, this formal construction of the convergent involves only symmetry-preserving matrix inversions and additions. The non-commutative matrix multiplications with the ${\bf a}_i \neq \mathds{1}$ would indeed break the  symmetries of the original operator.    

We must discuss here another technical aspect of the continued-fraction representation that requires careful treatment in the case of matrix-type coefficients. Coming back to the scalar case, and considering eq.~\ref{eqn:thiele4}, we observe that at  step   $(i)$ of the recursion process, the convergence of the continued fraction may be reached for a specific reference point $z_j$ for which the condition $f^{AC}(z_j)=f(z_j)$ has not been enforced yet ($j>i$). It happens whenever the corresponding propagated values $g_i(z_j)$ coincides with the propagated values $g_i(z_i)$ that serves as the stop condition for step (i). In such cases, we have thus that $g_i(z_j)^{-1} \to g_i(z_i)^{-1}$, leading explicitly to the condition $g_{i+1}(z_j)\to 0$:
\[
g_{i+1}(z_{j>i}) = \cfrac{g_i(z_j)^{-1}-g_i(z_i)^{-1}}{z_j-z_i} \to 0
\]
Iterating the recursion procedure another step, a "converged" partial denominators $g_{i+1}(z_{j})\to 0$ will produce a condition $g_{i+2}(z_j)\to\pm\infty$ for the next partial denominator value. For reference $f^{AC}(z_j)$ that are converged before being enforced, the continued expansion of the fraction is cut implicitly through the introduction of a pole in the subsequent partial denominators.

While such feature of the Thiele interpolation is harmless in the case of a scalar function, it becomes problematic when performing similar operations on matrices. In the case of a matrix, the convergence of the interpolant may occurs on a single rank-1 component of the whole matrix. Namely, the  ${\bf g}_{i+1}(z_j)$ matrix may become rank-deficient, corresponding to a situation where some directions for the reference tensor are converged while others are not. In practice, the difficulty occurs in the next step when attempting to generate the inverse of ${\bf g}_{i+1}(z_j)$. While this would be fine with infinite precision arithmetic, the finite dynamical range of the floating point number mantissa results in the pole masking all the other components contributions. This effectively stops the continued expansion for all matrix components, even the non-converged ones.    

As a simple mean to overcome this difficulty, we systematically use pseudo inverse during recursion. \red{In a nutshell, for a (close-to) singular matrix $A$, we write its Singular Value Decomposition (SVD) and the resulting pseudo inverse as:
\begin{align}
    A = U
    \left[\begin{array}{cc}
    \Sigma & 0 \\
    0  & \varepsilon
    \end{array}\right] V^{T},     \;\;\;  
     A^{\text{pinv}}\text{ (or } A^\oplus) = V
    \left[\begin{array}{cc}
    \Sigma^{-1} & 0 \\
    0  & 0
    \end{array}\right] U^{T},    
\end{align}
where $\Sigma$ and $\varepsilon$ are respectively the significant and negligible singular values of $A$, as revealed by the SVD decomposition. In this work, a ratio of $10^{-8}$ with  respect to the largest element of $\Sigma$ is chosen to filter out the small singular values that fall back in $\varepsilon$. }
Instead of propagating a pole to the subsequent partial denominators, the pseudo-inverse effectively cuts the corresponding rank-1 component in the recursion. Aside, from these considerations, the construction of the matrix valued continued fraction remains identical to the scalar one.

\subsection{Illustration: Resolvent of a simple Hamiltonian} 

\begin{figure}[h]
\begin{center}
 	\includegraphics[width=0.45\textwidth]{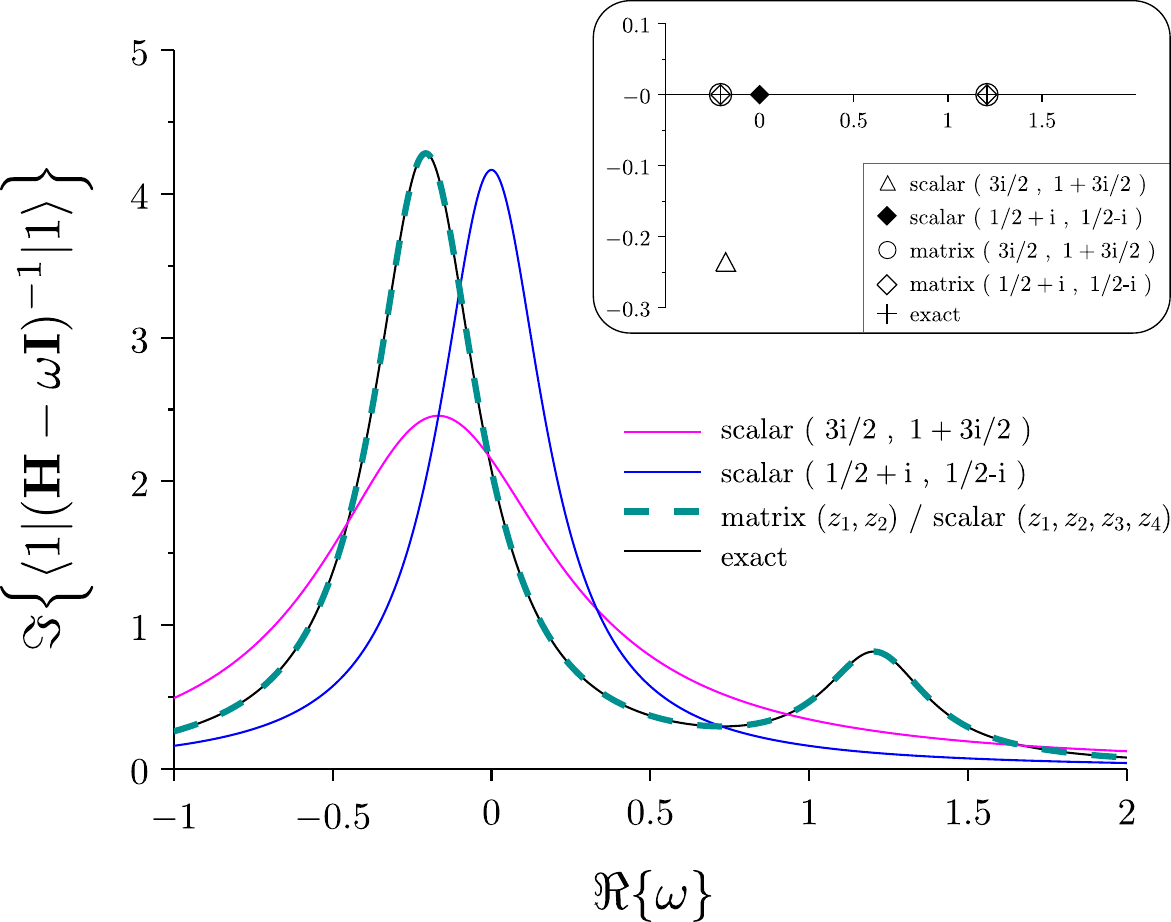}
    \end{center}
	\caption{ Representation in black of the reference $\rho_{11}(\omega+i\gamma)$ for the 2$\times$2 model   ($\gamma$=0.125). In pink and blue, the minimal model $\rho_{11}^{AC}(\omega+i\gamma)$ spectrum building a scalar continued-fraction representation with   $(z_1,z_2)=(3i/2,1 + 3i/2)$ and $(1/2 \pm i)$, respectively.
    The cyan dashed line shows that building a (2$\times$2) matrix-valued continued-fraction representation with  the same $(z_1,z_2)$ data points leads to an exact result. (Inset)  The position of the poles of the various continued-fraction representations is given.  }
	\label{fig:fig2x2model}
\end{figure}

To illustrate the above considerations, we start with a minimal two-level system  described by a 2$\times$2 Hamiltonian
\begin{equation}
     {\bf H} =\left[\begin{array}{cc}
    0 & 1/2 \\
    1/2  & 1
    \end{array}\right],
\end{equation}
yielding two eigenvalues at $(1\pm\sqrt{2})/2$ (about -0.207 and 1.207). We focus on the associated resolvent     ${\bf f}(z)=({\bf H}-z \mathds{I})^{-1}$. We plot in Fig.~\ref{fig:fig2x2model} the corresponding projected density of state $\rho_{11}(z)$:
\begin{equation}
\rho_{11}(z) = \Im\left\{ f_{11}(z) \right\}\;,\quad f_{11}(z) = \langle 1| \, ({\bf H}-z \mathds{I})^{-1} | 1 \rangle
\end{equation}
To set the reference, we calculate explicitly $\rho_{11}(\omega+i\gamma)$  with  $\gamma=0.125$  (black line). As expected, the associated spectrum is dominated by two  peaks centered on the eigenvalues of the Hamiltonian and a relative strength associated with the off-diagonal coupling magnitude.
To analyze the merits and difficulties of the AC technique, we compare now the corresponding continued fractions obtained from both scalar and matrix-valued continuations starting from the explicit calculation of the ${\bf f}(z)$ tensor for two-points sets $(z_1,z_2)$ of complex frequencies.

We first consider the continued fraction representation of the scalar quantity $f_{11}(z)$, expressed thus as follows:
$$
  f_{11}^{AC}(z) =   \frac{ 1}{b_1 + (z-z_1)/b_2}. 
$$
Here  $b_1$ and $b_2$ are complex scalars and $z_1$ is a complex energy at which $f_{11}(z_1)$ is explicitly calculated, yielding  $b_1 = 1/f_{11}(z_1)$. Calculating $f_{11}(z_2)$ for the second $z_2$ energy allows determining the termination factor $b_2$. From these two calculation, the functional form $f_{11}^{AC}$ is now fully determined and we can plot the corresponding projected density of state on the same ($\gamma=0.125$) energy line.

While there is an infinity of $(z_1,z_2)$ pairs, we focus on two typical cases: (i) considering first $(z_1,z_2)=(3i/2,1 + 3i/2)$, i.e. the real-part of $(z_1,z_2)$ equal to the diagonal entries of the Hamiltonian, 
the resulting $\rho_{11}^{AC}(\omega+i\gamma)$ is represented in Fig.~\ref{fig:fig2x2model}  (pink line). 
 As expected, this minimal resulting functional form can only capture one pole. The broad character of the peak suggests that in this case the $AC$ tends to average the two poles contribution as a single pole associated with a lifetime (i.e. located in the lower-half of the complex plane). The peak maximum is leaning close to the main contributing pole at -0.207. 
(ii) selecting now the ($z_1,z_2)=(1/2+i,1/2-i)$ symmetric sampling, the peak maximum corresponding now directly to the diagonal $\langle 1 |{\bf H}|1\rangle$ expectation value of the Hamiltonian. We verify thus that the peak energy location depends dramatically on the position of the chosen $(z_1,z_2)$. Further, choosing a complex symmetric $z_2 = z_1^*$ sampling enforces the location of the pole on the real axis,  so that the peak width is now controlled  by the $\gamma$ constant chosen when plotting $\alpha(\omega+i\gamma)$. To capture  2 poles with a scalar continued fraction, one must build a higher order functional form built from the knowledge of $f_{11}(z)$ calculated for at least four values of $z$.

We now consider the full (2$\times$2) tensor $\bf f$  and make again a minimal continued fraction representation:
\begin{align}
  {\bf f}^{AC}(z)  =  \left(    {\bf b}_1  + (z-z_1)  {\bf b}_2^{-1} \right)^{-1}  
\end{align}
where now the bold coefficients ${\bf b}_1$  and ${\bf b}_2$ are (2$\times$2) matrices. They are obtained from calculating  the full tensor $ {\bf f}(z)$  for the same $(z_1,z_2)$  sampling frequencies used before. As shown in Fig.~\ref{fig:fig2x2model}, the projection $\Im\{\langle 1 |  {\bf f}^{AC}(z) | 1 \rangle\}$ now reproduces perfectly the reference spectrum when plotted along the $(z=\omega+i\gamma)$ energy line (cyan dashed line). The resulting  $\rho_{11}^{AC}(z)$ is in fact independent of the choice of $(z_1,z_2)$ in this case. We note that the poles of $ {\bf f}^{AC}$ are canceling the determinant of the denominator ${\bf b}_1 + (z-z_1){\bf b}_2^{-1}$, which is of second order in z, accepting thus two poles.  

\red{It is worth noticing that the reconstruction of the full trace of $\rho(z)$ by adding the scalar continued-fraction representation of the diagonal matrix elements will yield the same number of poles as the one obtained by the matrix-valued continued-fraction representation. However, in our example, these poles will be located at the wrong energies. This should be put in perspective with the fact that obtaining the full resolvent tensor for a given (z)-frequency has the same cost as obtaining only the diagonal matrix-elements. Once again, it reveals the important fact that all matrix elements of the resolvent should share the same pole structure, a constraint which is naturally imposed by building a matrix-valued continued-fraction representation for the full tensor.}

To finish commenting on this simple but pedagogical example, we further plot in the Inset the position of the poles associated with the denominator of the continued fraction representation.  In the case of the scalar fit with $(z_1,z_2)=(3i/2,1 + 3i/2)$ (pink line), the pole is located in the second-quadrant (open upper triangle) with a large imaginary part. The model tries to account for the two separate peaks by a large broadened 
structure. The presence of such large structures with poles located away from the real-axis offers a useful diagnostic that the model fails resolving distinct features. Selecting symmetric (conjugate) sampling points $(z_1,z_2)=(1/2\pm i)$ leads to a peak corresponding to a single pole located on the real energy axis (black diamond). The resulting spectrum is not better than the broadened one in the sense that it yields even less weight on the secondary peak. We will discuss here below   the \textit{pros and cons} of symmetrizing the sampling $(z_k)$ complex energies to enforce strictly the complex conjugation property of a response function. In the other cases, scalar fit with 4 sampling frequencies coming by pairs of conjugated frequencies, or matrix-valued fits with 2 conjugated sampling frequencies, the poles are exactly located on the real axis at the exact eigenvalues of the model Hamiltonian.

\subsection{ Poles and residues of a continued fraction }

While the poles of a (2x2) matrix-valued continued-fraction   can be easily obtained as done above, we now generalize the search for the poles and associated residues for higher-order representations.
For the evaluation of matrix-valued continuous fractions, we use an adaptation of the well established fundamental recurrence formula: it can be shown that the continued fraction of $2n$-th order can be expressed as the ratio of two polynomials:
$$
  \cfrac{1}{ {\bf b}_1+} \; \cfrac{(z-z_1)}{{\bf b}_2 +} \; \cfrac{(z-z_2)}{  {\bf b}_3 +}  \cdots \cfrac{(z-z_{2n-1})}{  {\bf b}_{2n} } = 
 {\bf Q}^{-1}_{2n}(z) \cdot {\bf P}_{2n}(z)
$$
The numerators and denominators of the fraction's successive convergents are built iteratively starting from:
\begin{equation}
\begin{array}{lcl}
{\bf P}_{-1}(z) = \mathds{I}   & \quad\quad  &  {\bf P}_0(z) = 0\\
{\bf Q}_{-1}(z) = 0   &  \quad\quad  & {\bf Q}_0(z) = \mathds{I}.
\end{array}
\end{equation}
and then computing recursively 
\begin{align}
  {\bf P}_{k}(z) &=  {\bf b}_k \cdot {\bf P}_{k-1}(z) + (z-z_{k-1}){\bf P}_{k-2}(z) \\
  {\bf Q}_{k}(z) &=  {\bf b}_k \cdot {\bf Q}_{k-1}(z) + (z-z_{k-1}){\bf Q}_{k-2}(z)
\end{align}
The final resulting ${\bf P}_{2n}(z)$ and ${\bf Q}_{2n}(z)$ matrices can be either evaluated numerically for a specific $z$ value, or kept as polynomials in the variable z of order (2n-1) and (2n) respectively. 
This latter representation of ${\bf P}_{2n}(z)$ and ${\bf Q}_{2n}(z)$ allows in particular to obtain the poles $Z_{\lambda}$  of the continued fraction as the zeros of the ${\bf Q}_{2n}(z)$ polynomial, as well as the associated residues. In practice the poles can be extracted either as the eigenvalues of the ${\bf Q}_{2n}(z)$ polynomial companion matrix or directly from the finite eigenvalues of the following matrix pencil:~\cite{Celis_2024} 
\begin{equation}
    \left[\begin{array}{cccc}
    - {\bf b}_1 & z_1\mathds{I} & \\
    \mathds{I} & - {\bf b}_2 & z_2\mathds{I} \\
                & \mathds{I}     & \ddots  & \ddots\\
                &                 & \ddots  & -{\bf b}_{2n}
    \end{array}\right]\quad,\quad
    \left[\begin{array}{cccc}
    0 & \mathds{I} & \\
      &  \ddots & \ddots  \\
      &   & 0 & \mathds{I} \\
      &   &  & 0 
    \end{array}\right]
\end{equation}
Once the poles are known, the corresponding residues can also be obtained through linear algebra. To do so, we start by evaluating ${\bf Q}(z)$ value and its derivative ${\bf \dot{Q}}(z)$ at the desired pole frequency $Z_{\lambda}$, and diagonalize the corresponding pencil $\big\{{\bf Q}(Z_{\lambda}),{\bf \dot{Q}}(Z_{\lambda})\big\}$ in order to get a Taylor expansion of ${\bf Q}(z)$ around $Z_{\lambda}$:
\begin{equation}
{\bf Q}(Z_{\lambda}) = {\bf U} \cdot {\bf q} \cdot {\bf V}
\;,\quad
{\bf \dot{Q}}(Z_{\lambda})=\left.\cfrac{\dd {\bf Q}(z)}{\dd z}\right]_{Z_{\lambda}} = {\bf U} \cdot {\bf \dot{q}} \cdot {\bf V}
\end{equation}
\begin{equation}
{\bf Q}(Z_{\lambda}+\delta_z)\simeq {\bf U} \cdot \left( {\bf q} + \delta_z \,{\bf \dot{q}} \right) \cdot {\bf V} + \mathcal{O}(\delta_z^2)
\end{equation}
Here, the matrices ${\bf U}$ and ${\bf V}$ store the generalized left and right eigenvectors of the matrix pencil, while the diagonal matrices ${\bf q}$ and ${\bf \dot{q}}$ hold the corresponding generalized eigenvalues.
From this, the corresponding residue ${\bf R}_{\lambda}$ is obtained by using L'H\^{o}pital's rule on the 0-valued diagonal entries of ${\bf q}$
\begin{equation}
\begin{split}
{\bf R}_{\lambda} & = \lim_{z \rightarrow Z_{\lambda}} (z-Z_{\lambda}) \, {\bf Q}(z)^{-1}\cdot {\bf P}(z) \\    
& = {\bf V}^{-1}\cdot \left[ \lim_{\delta_z \rightarrow 0} \delta_z\,({\bf q} + \delta_z\,\dot{{\bf q}})^{-1}\right] \cdot {\bf U}^{-1} \cdot {\bf P}(Z_{\lambda})
\end{split}
\end{equation}
Such an identification of the poles and associated residues is often central to response functions analysis. For instance, in the case of the polarizability function, such residues are associated to the oscillator strengths. As a result, the continued fraction can be finally re-written as:
\begin{align}
 {\mathbf f}^{AC}(z) = \sum_{\lambda} \frac{{\bf R}_{\lambda}}{z-Z_{\lambda}}
\label{eqn:eqpoles}
\end{align}
providing thus directly useful insights about the spectral representation of the corresponding response function. 
 Such a capability makes Thiele's interpolation technique close in principle to the capabilities of integral methods that exploit quadrature rules to approximate the eigenvalues enclosed within a given contour. \cite{Asakura2009,Sakurai2010,Beyn2012,Polizzi2009,Polizzi2013,Saad2019}  

\subsection{ Symmetries }

The specific linear response functions that we will further investigate in the rest of the article, namely the Bethe-Salpeter polarizability tensor,  are even in the frequency space, namely:
\begin{equation}
     {\mathbf{f}}(\omega)= {\mathbf{f}}(-\omega)
\end{equation}
This implies that on general grounds, the extension to the complex plane of its spectral representation may be written:
\begin{align}
 {\mathbf{f}}(z) = \sum_{\lambda} \frac{{\bf R}_{\lambda}}{z-Z_{\lambda}} - \sum_{\lambda} \frac{{\bf R}_{\lambda}}{z+Z_{\lambda}} 
\label{eqn:eqpoles_sym}
\end{align}
Here, two strategies are possible: i) one can either simply include symmetric points in the sampling set, and calculate the associated response for, e.g., only the positive frequencies; or ii) we can remark that Eq.~\ref{eqn:eqpoles_sym} can be factorized:
\begin{align}
 \mathbf{f}(z) = 2\sum_{\lambda} \frac{Z_{\lambda} {\mathbf R}_{\lambda}}{z^2-Z_{\lambda}^2} 
\label{eqn:eqpoles_sym_fact}
\end{align}
and construct the symmetric Thiele continued fraction $\mathbf{f}^{AC}(y)$ on the set of $\big\{y_k=z_k^2\,,\;  \mathbf{f}(z_k)\big\}$ references. In practice, we found this last approach to be superior, especially when sampling frequency windows located in the high-frequency range, as explicit inclusion of the symmetric points would leads to rather uneven sampling distributions.

A second symmetry concerns conjugation,  namely the relation ${\bf f}(z^*)  = {\bf f}(z)^*$ of the response function. This property is associated with the fact that the response function is real-valued along the real-axis.
This symmetry relation can be exploited by adding to the $2n$ sampling complex frequencies $\{z_k\}$ located in the upper complex half-plane, the corresponding $\{z_k^*\}$ frequencies situated in the lower quadrants. This doubles at no cost the number of explicitly calculated response functions used to build the continued-fraction representation, allowing to double the recursion depth and the number of poles that can be captured along the real axis. An important consequence of such an approach is that the poles of the continued-fraction are more likely to be real-valued. 

The use of this complex-conjugation symmetry may however not be advisable when dealing with resonances made out of a quasi-continuum of discrete excitations or poles, namely a branch cut. In that case, we do not wish to reconstruct all individual poles with energies located on the real-energy axis, but only their resulting broad envelope. Such an envelope is associated with discrete poles in the lower complex half-plane, that is with a negative imaginary part related to the envelope peaks half-width. These complex energy poles associated to resonances are often physically very meaningful,~\cite{KATO1998}  such as in the case diluted defects,~\cite{LEVITT2023}  or collective modes, as illustrated with the specific example  of plasmons here below. 

\subsection{Response function sampling}

A striking advantage of resolvent-based techniques is that one can select freely both  the frequency range as well as the projected subspace that is being sampled, while retaining the entire contribution from all degrees of freedom of the underlying operator: in the rest of this study, we will focus on the $3\times3$ dipolar components of the BSE polarizability tensor, which reflect in turn the contributions coming from the entire BSE eigen-spectrum.  

\begin{figure}[t]
  \begin{center}
  	\includegraphics[width=0.5\textwidth]{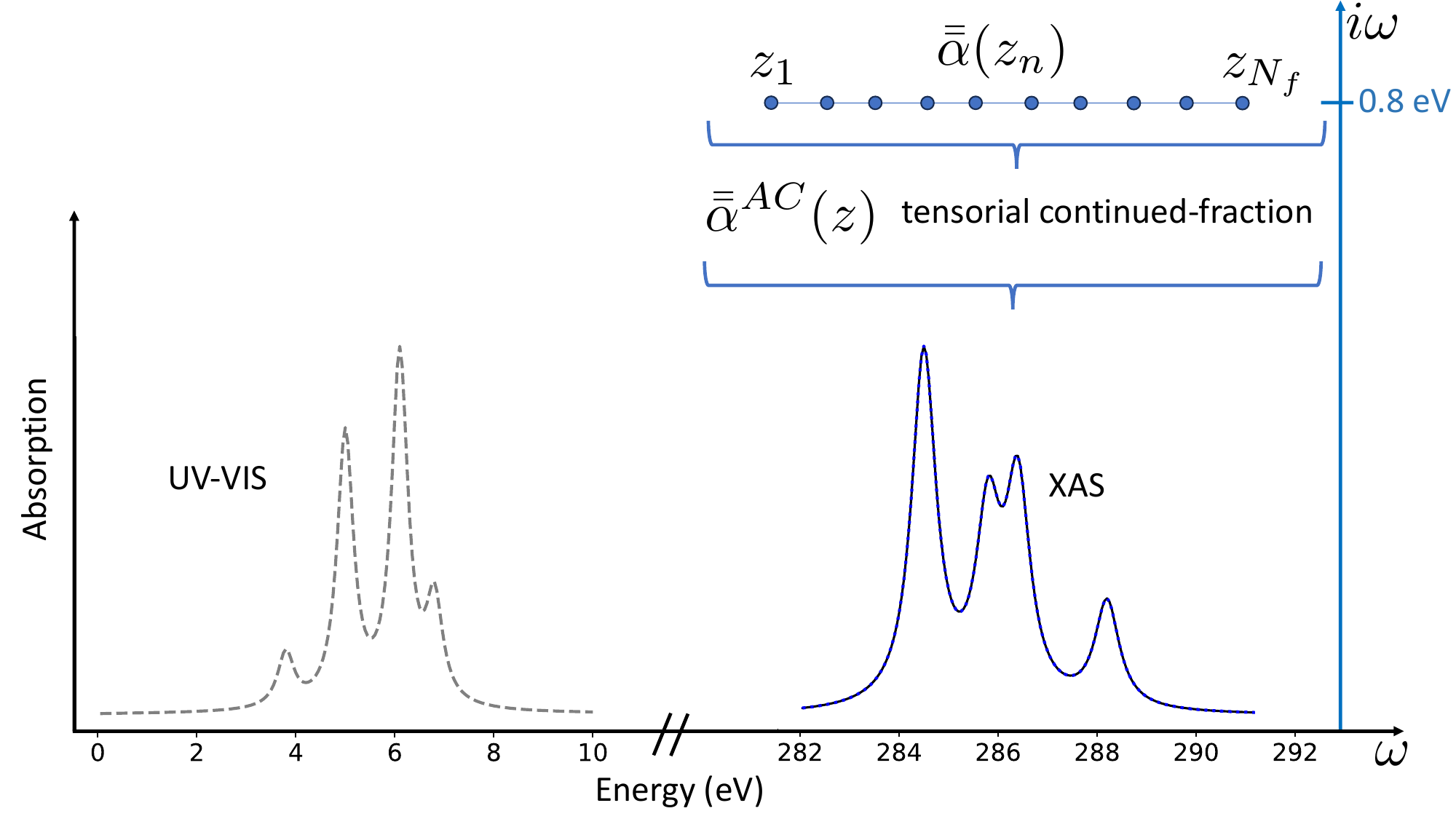}
  \end{center}
  \caption{ Schematic representation of the C$_{60}$ fullerene valence (dashed) and core (full thick line) absorption spectrum. The calculation of the (3$\times$3) polarizability tensor $\bar{\bar\alpha}(z_k)$ on a very coarse grid of frequencies $\lbrace z_k \rbrace$ in the complex plane allows constructing a tensorial continued-fraction representation $\bar{\bar\alpha}^{AC}(z)$ that captures accurately the absorption spectrum close to the real-frequency axis in a selected energy range.  The position of the selected sampling $\lbrace z_k \rbrace$ complex energies, as represented here, allows reconstructing selectively the X-ray absorption spectrum at high energy. For  response functions, the number $N_f$ of sampling frequencies $(z_k)$ must be even.    }
  \label{fig:figtoc}
\end{figure}

While there are certainly a large number of ways of selecting the representative $\lbrace z_k \rbrace$ frequencies in the complex plane, we will consider in the following simple regular grids, $z_k = \omega_0 + k \Delta\omega + i\Gamma$, with $k=1 \rightarrow N_f$ and $N_f$  the  even  number of sampling points. 
In order to demonstrate the efficiency of this simple approach, without any \textit{a priori} knowledge of the exact number and energy position of the poles of the response function, we will consider $\Gamma$=0.8~eV and $\Gamma$=0.4~eV sampling height, together with a fixed spacing $\Delta\omega \simeq \Gamma / 1.5$. 
Such settings stems from a previous work focusing on obtaining the $GW$ Green's function spectral representation ${A}^{GW}(\omega)$ along the real axis by means of analytic continuation, calculating the $GW$ Green's function $G(z)$ over a few $(z_k)$ frequencies in the complex plane.\cite{Duchemin2020,Duchemin2025} Concerning the specific case of BSE calculations, we will show that the continued fraction representation built from these limited data points offers an accurate representation of the Bethe-Salpeter (3$\times$3) polarizability tensor $\bar{\bar \alpha}(\omega+i\gamma)$  in the $\omega \in [\omega_0, \omega_0 + N_f \Delta\omega ]$ energy range along the real axis, where $\gamma=$~0.2eV is  a realistic small experimental broadening. 
This strategy is schematically represented in Fig.~\ref{fig:figtoc} in the case of C$_{60}$. For sake of visibility, we switch to the ``{d}ouble-bar" notation for the polarizability tensor.

A last point to be discussed   is the order we use for the inclusion of the sampling frequencies during the interpolant construction, as it is known  that this order affects the continued fraction numerical stability. While a greedy selection of the frequencies together with an early termination condition has proven to bring reliable results,\cite{Celis_2024} we chose here the slightly more computationally demanding approach that consists in determining at each step the frequency that minimizes the overall error of the interpolant over the remaining sampling set:
\begin{equation}
    \sum _ {k>j} |f^{AC}_j(z_k) - f(z_k)|^2.
\end{equation}
This strategy proved to bring particularly stable results and is motivated by the very minimal computational effort involved in the continued fraction coefficient construction with respect to the explicit calculation of the response function for all the sampling frequencies.

\section{ Application to the Bethe-Salpeter polarizability tensor }

 While presenting here above the matrix-valued analytic continuation approach to response functions, we now specialize to the Bethe-Salpeter polarizability tensor.     Details about the many-body Green's function $GW$ \cite{Hedin1965,Farid99,Str80,Ary98,ReiningBook,Ping2013,Golze2019,Bruneval2021,Marie2025} and BSE \cite{Csanak1971,Strinati1984,Blase2020,Holzer2025} perturbation theories can be found in thorough reviews and books. We focus here below on the expression of the Bethe-Salpeter 4-point susceptibility $L$ as the  resolvant of the BSE two-body Hamiltonian, together with the expression of the molecular polarizability tensor $\bar{\bar{\alpha}}^{BSE}$ as a function of   $L$.  More details  can be found e.g. in Ref.~\citenum{Ljungberg2015}. We finally discuss how to obtain iteratively the needed few $\bar{\bar{\alpha}}(z_k)$ reference tensors to set-up a matrix-valued representation $\bar{\bar{\alpha}}^{AC}(z)$ as described above.

\subsection{The Bethe-Salpeter resolvent and the polarizability}

 We start directly from the relation between the Bethe-Salpeter four-point response function $L$, namely  the derivative of the one-body Green's function by an external non-local perturbation, and the Bethe-Salpeter two-body (electron-hole) Hamiltonian $\mathcal{H}^{BSE}$:
\begin{align}
    L(\omega) = &  \left( \omega \mathds{I} - \mathcal{H}^{BSE}\right)^{-1} \Delta  \label{eqn:LtoH} \\ 
              = &  \left( \omega \Delta - \Delta\mathcal{H}^{BSE}\right)^{-1} \label{eqn:LtoH_2}
\end{align} 
expressed in the transition space between molecular orbitals (MOs) where the $\mathcal{H}^{BSE}$ Hamiltonian and $\Delta$ operators read:
 \begin{align}
   \mathcal{H}^{BSE} = 
   \begin{pmatrix}
A & B\\
-B^* & -A^*
\end{pmatrix}
\;,\quad
   \Delta = 
   \begin{pmatrix}
\mathds{I} & \\
 & -\mathds{I}
\end{pmatrix}
\end{align}
The 4-point susceptibility $L$ is thus linked to the resolvent of the BSE Hamiltonian through a multiplication with $\Delta$  that reflects the specific metric associated with resonant/non-resonant channels. 
Within the adiabatic BSE/$GW$ approach, the $A$ and $B$ matrix elements are expressed over the (occupied)$\times$(virtual) transition product-space as:  
\begin{align}
    A_{ia,jb} &= ( \varepsilon_a^{QP} - \varepsilon_i^{QP}) \delta_{ij} \delta_{ab} + {\kappa}(ia|jb) -W_{ij,ab}  \label{eq:Amat}\\
    B_{ia,jb} &=  \kappa(ia|jb) -W_{ib,aj}     \label{eq:Bmat}
\end{align}
where the (i,j) and (a,b) indices point here to occupied and unoccupied energy levels, respectively.  It is worth mentioning here that these coefficients are never explicitly calculated, as the size of the $\mathcal{H}^{BSE}$ Hamiltonian grows too rapidly with that of the system. Instead we make use of the underlying 2-point structure of the different terms involved in the Bethe-Salpeter kernel, in conjunction with (Coulomb-fitting) resolution-of-the-identity techniques, \cite{Vahtras1993,Ren2012,Duchemin2017} to obtain efficiently the action of $\mathcal{H}^{BSE}$ on the BSE vectors.  The prefactor $\kappa$ equals 2/0 for singlet/triplet excitations. The $\lbrace \varepsilon_{a/i}^{QP} \rbrace$ are the $GW$ quasiparticle energies.
The statically (adiabatic) screened Coulomb matrix elements are defined as:
\begin{align}
    W_{ij,ab} &= \int d{\bf r}d{\bf r}' \; \phi_i({\bf r}) \phi_j({\bf r}) W({\bf r},{\bf r}'; \omega=0) \phi_a({\bf r}') \phi_b({\bf r}')
    \label{eq:wiajb}
\end{align}
where we take real molecular orbitals (MO) for finite size systems.   Such a formulation resembles that of Casida for TD-DFT, replacing the Kohn-Sham eigenvalues by the $GW$ quasiparticle energies, and the $f_{XC}$ TD-DFT kernel by the nonlocal and screened electron-hole interaction (-$W$).  As such, once the $GW$ outputs available, the BSE calculations can be performed at the same cost as standard TD-DFT. 
We will not discuss here the recent progresses in low-scaling $GW$ calculations. \cite{Rojas1995,Foerster2011,Neuhauser2014,Liu2016,Vlcek2017,Wilhelm2018,Kim2020,Forster2020,Kutepov2020,Duchemin2021,Forster2021,Tolle2024}

We target in the present study the Bethe-Salpeter polarizability tensor:
\begin{align}
     \alpha_{\mu\nu}(\omega) = - \langle  D_{\mu} | L(\omega) | D_{\nu} \rangle
     \label{eqn:LtoAlpha}
\end{align}
where  $D_{\mu/\nu}$ are the dipole transition vector in the product-space of occupied   to unoccupied MOs.  The operator $\hat x_{\mu}$ is the  position operator in the ($x_{\mu}$) direction, and $D_{\mu}^{ia}=D_{\mu}^{ai}=\langle \phi_i \phi_a |  x_{\mu} \rangle$.  Clearly, a similar approach could be used for any shape of the external perturbation, beyond the simple potential ramp.  We take the MOs $\lbrace \phi_i,\phi_a\rbrace$  to be real-valued in the case of finite size systems. 
The polarizability allows obtaining the optical absorption spectrum:
\begin{align}
    \sigma(\omega) = \frac{4\pi \omega}{3c} \text{Tr}\left[ \Im \bar{\bar \alpha}(\omega)  \right] 
    \label{eqn:absorption}
\end{align}
with $\text{Tr}$ the trace operator. Adding some imaginary part ($\gamma$) to the targeted frequency ($\omega \rightarrow z=\omega+i\gamma$) allows mimicking experimental broadening. In what follows, we will adopt the notation (z) to design a complex frequency.

\subsection{Iterative calculation of $| L(z)D_{\mu} \rangle$}

Following eq.~\ref{eqn:LtoH}, the traditional approach to BSE consists in  searching the poles of $L$ as the eigenvalues of the BSE Hamiltonian problem:
\begin{align}
     \mathcal{H}^{BSE}  | \psi^{BSE}_{\lambda} \rangle = \Omega_{\lambda} | \psi^{BSE}_{\lambda} \rangle
\end{align}
where the BSE 2-body (electron-hole) eigenstates read:
\begin{align}
    \psi^{BSE}_{\lambda}({\bf r}_e,{\bf r}_h) = \sum_{ia} \left[ X_{ia} \phi_i({\bf r}_e) \phi_a({\bf r}_h) + Y_{ia} \phi_i({\bf r}_h) \phi_a({\bf r}_e) \right]
\end{align}
The knowledge of the BSE eigenstates is a very valuable asset that permits analyzing the physical nature of the excitations. As an alternative approach, the vectors
$| L(z)D_{\mu} \rangle$  can be directly obtained as the solution of the linear system using the symmetric form \ref{eqn:LtoH_2}:
\begin{align}
    L(z)^{-1} \,\left| L(z)D_{\mu} \right\rangle = \left|D_{\mu}\right\rangle 
\end{align}
namely
\begin{align}
    \left( z\Delta - \Delta\mathcal{H}^{BSE} \right) \,\left| L(z)D_{\mu} \right\rangle = \left|D_{\mu}\right\rangle 
\end{align}
where ($x_{\mu}$) spans the 3 directions of space. 
In this case, the linear response vectors $| L(z)D_{\mu} \rangle$ can be obtained through iterative techniques such as the generalized minimal residual (GMRES) approach. \cite{Saad1986} 
Once the $| L(z)D_{\mu} \rangle$ vectors are known, the BSE polarizability tensor can be built straightforwardly by taking the scalar products $\langle D_{\nu} |   L(z)D_{\mu} \rangle$.   \red{As such, calculating the full $\bar{\bar \alpha}(\omega)$ tensor can be done with the same cost as calculating only its diagonal matrix elements. }
Such a technique was used in Ref.~\citenum{Kehry2023}  to calculate the absorption spectrum in  specific energy ranges  sampled with fine frequency grids $z_k = k\Delta \omega + i\gamma$ where $\gamma$ represents a typical experimental broadening. Values of  $\Delta \omega \simeq 0.05$ eV and   $\gamma \simeq 0.124$ eV (100 cm$^{-1}$) were used for a very detailed scan of the absorption in the typical X-ray absorption range associated with the K-edge of specific elements belonging to small molecules.  

As noted previously, \cite{Kehry2020} the present approach allows  dealing with dynamical (frequency-dependent) Hamiltonians. Recent studies \cite{Rohlfing2000,Ma2009,Loos2020,Bintrim2022,Monino2023} have paved the way to going beyond the adiabatic approximation at the BSE level. 
This stands as another potentially interesting advantage over standard iterative eigenvalue solver approaches. Further, the GMRES scheme allows preconditioning as we now discuss. 

\subsection{ Preconditioning }

To achieve numerical efficiency, we rely on the left preconditioned version of the initial $L(z)^{-1} X(z)  = b$ linear system: $\widetilde{L}\,\big(L^{-1} X -b\big)$.
In this context, the product $\widetilde{L}L^{-1}$ should thus be as close as possible to the identity operator, and the application of $\widetilde{L}$ should also be performed efficiently. As a both physically and computationally motivated approximation to the four point BSE response function $L(z)$, we use as a preconditioner the RPA susceptibility $\widetilde{L}(z)=L^{RPA}(z)$, that can be obtained from the 4-points free electron susceptibility $L_0(z)$ through the following Dyson equation:
\begin{align}
    L^{RPA}(1,2&; 1',2') = L_0(1,2;1',2') + \int d3456 \; L_0( 1,4;1',3) \nonumber \\ 
&\times  \; \delta(34) v(3,6) \delta(56) \; \times \; L^{RPA}(6,2;5,2')  
\label{eq:LDysonRPA}
 \end{align}
However, we do not wish to invert the Dyson equation to compute explicitly the 4-points $L^{RPA}(z)$, but simply need its application to the residues obtained at each iteration of our optimization scheme. In order to do so, we rewrite it as:
\begin{align}
    L^{RPA}(1,2&; 1',2') = L_0(1,2;1',2') + \int d3456 \; L_0( 1,4;1',3) \nonumber  \\ 
&\times \; \delta(34)W(3,6)\delta(56) \;  \times \;L_0(6,2;5,2') 
\label{eq:LDysonRPAbis}
 \end{align}
  In the frequency and electron-hole product space, $L_0$ is diagonal:  
\begin{equation}
\begin{split}
 [L_0(z)]_{ia,jb} = & \phantom{-} \frac{\delta_{ab} \delta_{ij}}{z -(\varepsilon_a-\varepsilon_i)+i\eta} \\
 [L_0(z)]_{ai,bj} = & - \frac{\delta_{ab}\delta_{ij}}{ z + (\varepsilon_a-\varepsilon_i)-i\eta} 
\end{split}
\end{equation}
where the positive infinitesimal can be omitted for complex frequencies.  This diagonality of $L_0$ in the electron-hole product space insures that the products with $L_0$ in eq.~\ref{eq:LDysonRPAbis} can be performed swiftly directly on the response vector coefficients, while the action of the  $\delta(\br,\br')W(\br,\br'';z)\delta(\br'',\br''')$  kernel is now applied in the same manner as the $(ia|jb)$ term of equation \ref{eq:Amat}, replacing the bare coulomb interaction $v(\br,\br')$ by its screened counterpart  $W(\br,\br'; z)$  taken at the desired $z$ complex frequency.


\section{Results}

\subsection{Validation for valence excitations}

Low-lying valence excitations can be obtained efficiently with standard iterative eigenvalue solvers, providing both the lowest BSE eigenvalues and the associated eigenstates. In that case, techniques  extracting the absorption spectrum directly  may be considered as less interesting  since the corresponding eigenvectors are not available. Within the scope of the present paper, the study of low-energy excitations allows primarily comparing directly the results of our analytic continuation scheme with respect to reference BSE calculations. We however consider two important systems for which the present approach proves interesting even in the case of  valence  optical excitations. The first system is the C$_{60}$ fullerene, for which the main absorption peaks are buried in a very large number of dark transitions, and the second is the Ag$_{20}$ metallic cluster presenting a resonance. In the first case, obtaining the important peaks with standard techniques may require calculating hundreds of dark states before reaching the relevant absorption energies. In the second case, the resonance  may be composed of an increasingly large number of densely packed individual excitations in the large metallic cluster size limit, merging into an  ``{e}nveloppe" that is the structure of interest.  In that respect, the present analytic continuation approach proves very efficient since allowing to reveal the main absorption features with few reference calculations. As an intermediate system of interest, we also consider the [6,6]-phenyl-C61-butyric acid methyl ester (PCBM) fullerene derivative  with the explosion of the well-resolved absorption lines of C$_{60}$ into broad peaks due to symmetry breaking by functionalization. Before describing these paradigmatic systems, we start with a small but complex molecule, the $\beta$-dipeptide, to illustrate again, but on a realistic system, the interest of building a tensorial continued fraction representation. 

\subsubsection{ The $\beta$-dipeptide }

\begin{figure}[t]
\begin{center}
 	\includegraphics[width=0.49\textwidth]{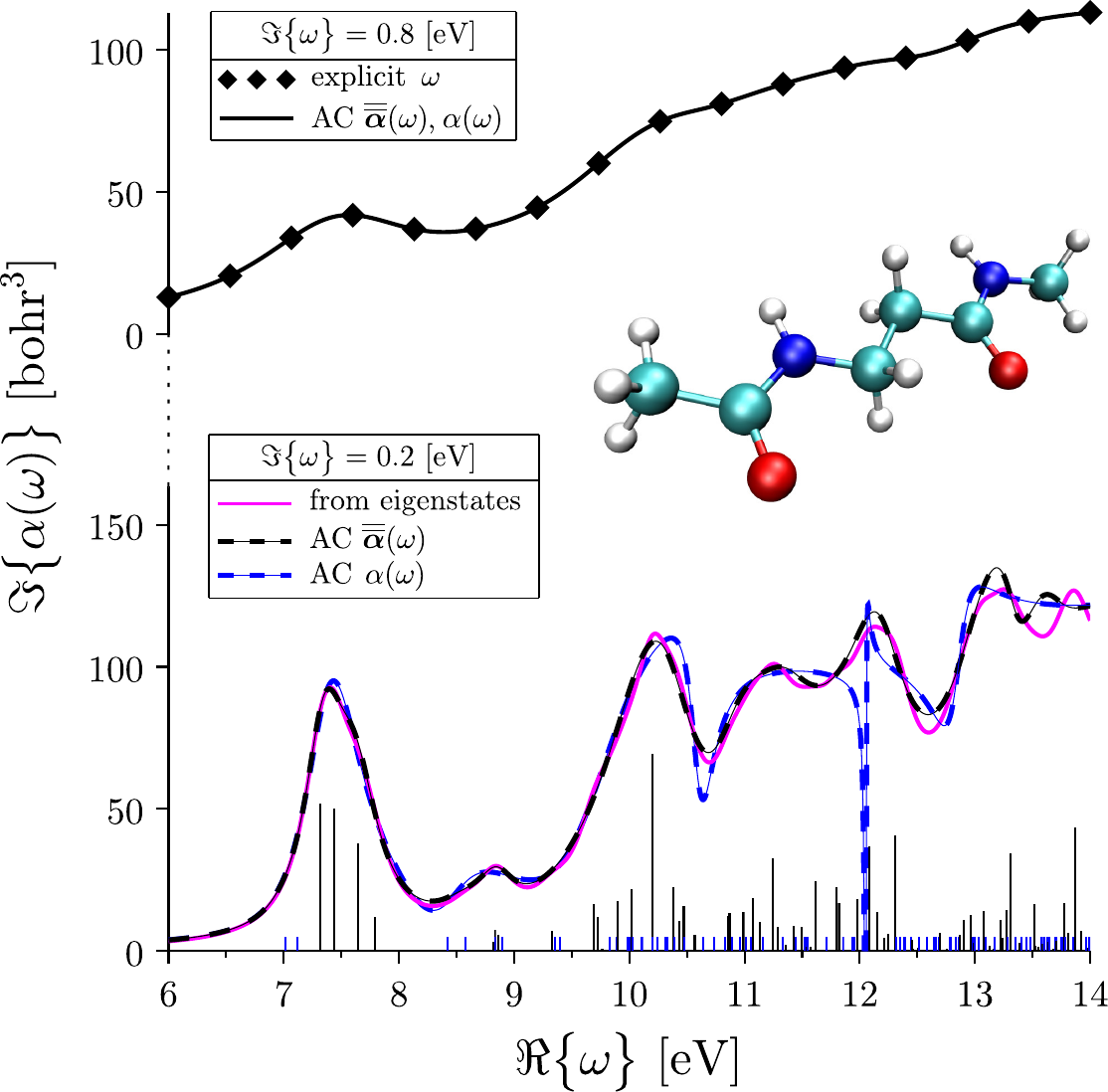}
    \end{center}
	\caption{ Imaginary part of the averaged trace of the polarizability tensor (labeled simply $\Im\alpha$) for the $\beta$-dipeptide (see Inset). The vertical bars are associated with the energy position and oscillator strength of the explicit BSE bright (black) and dark (short blue) eigenstates. The pink full curve is the reconstruction of $\Im\alpha(\omega+i\gamma)$   from the BSE eigenstates with $\gamma$=0.2~eV. The black diamonds are the $\Im\alpha(z_k)$ values ($z_k = \omega_k + i\Gamma$)   from the explicitly calculated $\bar{\bar{\alpha}}(z_k)$ tensors with $\Gamma$=0.8~eV. Based on the related tensorial continued-fraction representation $\bar{\bar \alpha}^{AC}(z)$,  $\Im\alpha^{AC}(z)$ is plotted along the $(z=\omega+i\Gamma)$ (upper panel, full black) and $(z=\omega+i\gamma)$ (lower panel, dashed black) energy axes. The blue dashed curve represents $\Im\alpha^{AC}(\omega+i\gamma)$ but with a scalar continued-fraction construction of the averaged trace.  }
	\label{fig:figdipeptide}
\end{figure}

The $\beta$-dipeptide  was part of several benchmarks due in particular to its intramolecular donor-acceptor character. \cite{Serrano1998,Casanova2019,Loos2021}  Even though of medium size, this molecule presents a rather complex low-energy spectrum with a large number of distinct $\pi$-$\pi^*$, $n$-$\pi^*$ and Rydberg excitations with a large distribution of oscillator strengths due to a mixed Frenkel and intramolecular charge-transfer character (see e.g. Supporting Information Ref.~\citenum{Loos2021} where the molecule atomic coordinates are given). As such, the $\beta$-dipeptide stands as a difficult test case for the present scheme. Our calculations are performed at the BSE/ev$GW$@PBE0 cc-pVTZ level. \red{Input Kohn-Shan eigenstates are generated with the {\sc{Orca}} package \cite{Neese2022,Neese2023} for all examples treated in the present study. } A ball-and-stick representation of the dipeptide is given as an Inset in Fig.~\ref{fig:figdipeptide}.

We start by calculating explicitly  200 BSE eigenstates that  are represented in Fig.~\ref{fig:figdipeptide}  with vertical  black bars proportional in length to the oscillator strength for bright excitations. Dark excitations are represented by small blue ticks of arbitrary length. 
From these BSE eigenstates, we reconstruct the imaginary part of the averaged trace of the  polarizability tensor:
\begin{align}
   \frac{1}{3} \Im Tr\; \bar{\bar \alpha}(\omega+i\gamma)  = \Im \frac{2}{3} \sum_{\lambda} \frac{ \Omega_{\lambda} \left[ \sum_{\mu = x,y,z} (d_{\mu}^{\lambda})^2  \right]  }{  \Omega_{\lambda}^2 - (\omega+i\gamma)^2 } 
   \label{eqn:avtrace}
\end{align}
including a small $\gamma$=0.2~eV broadening to mimic experimental broadening. Such a quantity, that we label simply $\Im \alpha(\omega+i\gamma)$, is directly related to the optical absorption spectrum through eq.~\ref{eqn:absorption}. The $\Omega_{\lambda}$ are the BSE eigenstates energies associated with the $  d_{\mu}^{\lambda} $ BSE transition moments, with e.g. $d_x^{\lambda} = \sqrt{2} \sum_{ia} \langle \phi_i | \hat{x} | \phi_a \rangle (X_{ia}^{\lambda} + Y_{ia}^{\lambda})$. We plot (pink line) $\Im \alpha(\omega+i\gamma)$ for $\omega$ up to 14~eV (the 200th BSE eigenstate has an energy of about 15~eV). 

We now set up our computational protocol, calculating as described above the polarizability tensor $\bar{\bar{\alpha}}(z)$ for a very-coarse grid of complex frequencies $(z_k = \omega_0 + k\Delta\omega +i\Gamma)$ with $\Gamma$=0.8 eV and  $\Gamma=1.5\times\Delta\omega$ . This scheme will be labeled AC(0.8) in what follows. The corresponding $\Delta\omega \simeq 0.53$ eV spacing yields 16  complex frequencies with a real-part in the [6,14]~eV energy range. As such, only 16 $\bar{\bar \alpha}(z)$ tensors are calculated iteratively. The corresponding $\Im \alpha(z_k)$ values are represented with black diamonds in Fig.~\ref{fig:figdipeptide}. From these data points, we set-up a tensorial continued-fraction representation $\bar{\bar \alpha}^{AC}(z)$.   We plot (black line, upper panel) the corresponding averaged trace $\Im\alpha^{AC}(\omega+i\Gamma)$ along the  $(z=\omega+i\Gamma)$ energy line. By construction of the continued-fraction functional form, this curve goes exactly through the reference black diamond. 

We further use the $\bar{\bar \alpha}^{AC}(z)$ functional form to plot $\Im\alpha^{AC}(\omega+i\gamma)$ with $\gamma$=0.2~eV (dashed black line, lower half).  Even though based on few reference points, the tensorial continued fraction representation captures nicely the distribution of spectral weight as given by the reference pink curve. The agreement is excellent at the bottom of the spectrum, degrading only in the upper spectrum.    We can speculate that data  points on the high energy-side of the energy window includes contribution from BSE poles located at higher energy.   As discussed above, the tensorial continued-fraction functional form with 16 sampling frequencies can only capture 3$\times$16/2=24 poles, that is much less than the number of bright excitations in the chosen energy range. Still, the main structures are well captured. 

We now build directly a \textit{scalar} continued-fraction representation of the polarizability tensor averaged trace
from the calculated $(\sum_{\mu=1,3}  \alpha_{\mu\mu})(z_k)/3$ values. The imaginary part of this obtained functional form is reproduced along the $(\omega+i\gamma)$ energy line (blue dashed). Clearly, while the agreement remains good in the low energy range, the agreement with the reference spectrum degrades at higher energy. An instability     appears at $\sim$12.1 eV associated with a pole in the first quadrant. As shown further in the  Supplementary Material, the agreement degrades even more significantly when attempting to fit separately specific $\alpha_{\nu\nu}(z_k)$ components of the polarizability tensor. 

The present example illustrates on a realistic system that the tensorial, or matrix-valued, continued-fraction   approach is much more stable. This is what we will be now using in the following. Concerning the quality of the results, the reported calculations illustrate that the distribution of spectral weight is nicely reproduced at low cost provided that one does not wish to resolve closely lying structures. This could be certainly obtained by zooming on a specific energy range, namely accumulating reference points closer to the real axis and in a restricted energy range, but at the cost of an increasing  number of calculations. 

\subsubsection{Gas phase C$_{60}$ fullerene}

We now consider the case of the C$_{60}$ fullerene. From superconducting fullerides to its use as a standard acceptor in organic solar cells,  pristine C$_{60}$,   and its derivatives, have been widely studied in relation with their unique properties. Gas phase experimental data (for a review, see Ref.~\citenum{Smith1996}) reveal dominant valence absorption peaks with energies at  3.8 eV, 4.8-4.9 eV, 5.95-6.05 eV in the UV-visible energy range.     The experimental absorption cross section indicates that the  structure at about 6 eV is the strongest, followed by the one at 5 eV, while the  band at 3.8 eV is significantly weaker. The absorption energies and cross sections slightly depend on the experimental conditions, with temperature ranging typically between 600~K and 700~K as needed to vaporize the C$_{60}$ samples.

\begin{figure}[t]
\begin{center}
 	\includegraphics[width=0.49\textwidth]{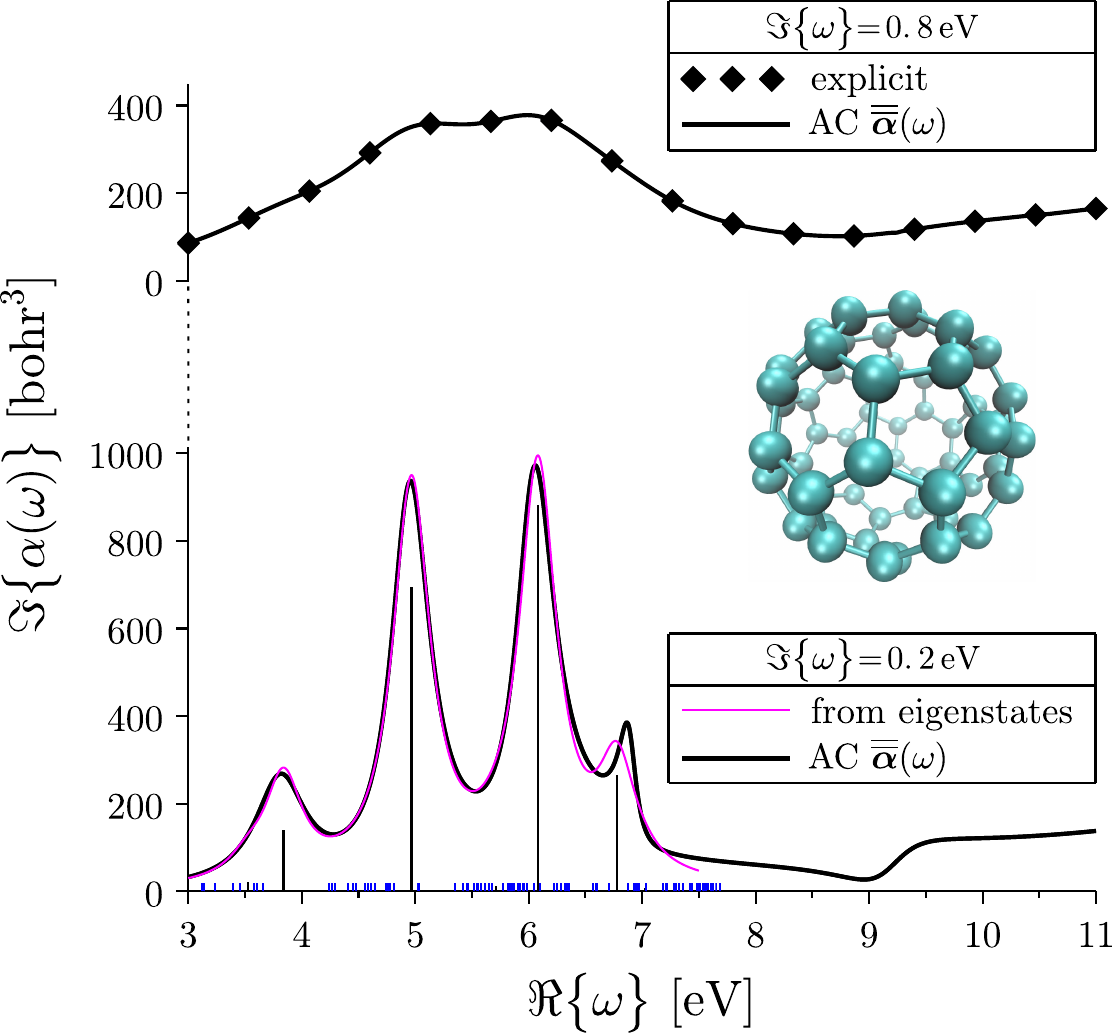}
    \end{center}
	\caption{ Same as  Fig.~\ref{fig:figdipeptide} but for C$_{60}$. Full black line lower panel:  plot of $\Im\alpha^{AC}(\omega+i\gamma)$ with $\gamma$=0.2 eV  within   the AC(0.8) scheme. We do not study here the results of the scalar continued-fraction construction.  }
	\label{fig:figc60}
\end{figure}

We provide in Fig.~\ref{fig:figc60} a plot similar to Fig.~\ref{fig:figdipeptide} but for the fullerene.  Calculations are performed at the BSE/ev$GW$@PBE0 cc-pVTZ level on the B3LYP cc-pVTZ geometry  \red{ (geometry provided in the SM)}.  We calculate  420 singlet BSE eigenstates,  with energy up to  7.7~eV. Out of these 420 eigenstates, only four (3-fold)  excitations are bright, with energies 3.839 eV (f=0.163), 4.967~eV (f=0.813), 6.082~eV (f=1.030)  and 6.780~eV (f=0.309) as compiled in the last line (Eigen.) of Table~\ref{tab:tableC60}. We give 3 decimals for sake of comparison later below. These theoretical data are in good agreement with experiment for the 3 lowest absorption lines concerning both the energies and relative oscillator strengths. 

The dramatically large proportion of dark states is associated with the fullerene high symmetry (I$_h$ icosahedral group). For sake of indication, the lowest bright excitation manifold is composed of the 55$^{th}$-57$^{th}$ eigenstates, while the last manifold at 6.78~eV corresponds  to the 299$^{th}$-301$^{st}$   BSE eigenstates. 
As such, capturing the ``low energy" excitation spectrum already requires calculating hundreds of BSE eigenstates with the standard eigenvalue solver approach. 
We did not seek to capture  higher energy features using the explicit search for the  BSE eigenstates.   From these calculated BSE eigenstates, we reconstruct (pink line, Fig.~\ref{fig:figc60})  the averaged trace of the imaginary part of the  polarizability tensor (see eq.~\ref{eqn:avtrace}) that we plot along the $(\omega+i\gamma)$ energy axis with $\gamma$=0.2~eV. This will serve again as the reference spectrum.

We now use the AC(0.8) scheme. Namely, we calculate explicitly $\bar{\bar \alpha}(z_k)$ tensors with $(z_k =  \omega_0 + k\Delta\omega +i\Gamma)$ and $\Gamma=1.5\Delta\omega=0.8$~eV (see corresponding black diamonds Fig.~\ref{fig:figc60}). This represents 16 reference ($z_k$) frequencies with a real part in the $[3,11]$~eV energy range. From these data points, we build a matrix-valued continued-fraction   $\bar{\bar \alpha}^{AC}(z)$ functional form that we use to plot the imaginary part of the averaged trace ${\Im} \alpha(z)$ along the $(\omega+i\Gamma)$  and $(\omega+i\gamma)$ energy axes with $\Gamma$=0.8~eV and $\gamma$=0.2~eV (upper and lower panel black full lines, respectively). We choose here a larger energy range (as compared to the available BSE eigenstates) to exemplify that larger portions of the spectrum can be obtained at  low cost. 

\begin{table}[t]
\begin{center} 
\begin{tabular}{ c|cccc} 
                  &  S$_{55-57}$        & S$_{197-199}$       & S$_{236-238}$       & S$_{299-301}$   \\
  AC(0.8)         & 3.813  (0.227) & 4.948   (0.837)     &  6.038   (0.982)    & 6.890  (0.132)\\
\;\;+$\alpha(z^*)$ & 3.839 (0.162) & 4.966 (0.813) & 6.080 (1.028) &  6.781 (0.313) \\
   AC(0.4)         & 3.839 (0.162) & 4.966 (0.814) & 6.079 (1.033) & 6.773  (0.316)  \\
    Eigen.        & 3.839 (0.163) & 4.967 (0.813) & 6.082 (1.030) & 6.780 (0.309)  \\ 
\end{tabular} 
\end{center}
\caption{ Bethe-Salpeter lowest singlet bright excitations in $C_{60}$ at the BSE/ev$GW$@PBE0 \red{cc-pVTZ} level (B3LYP geometry). Energies are in eV (oscillator strength in parenthesis). The notation e.g. S$_{55-57}$ indicates that we report on the degenerate 55th-to-57th excitations. We compare the results of standard BSE eigenstates   calculations (Eigen.) to that obtained with analytic continuation using the AC(0.4) scheme  and the rough  AC(0.8)  alternative (see text). The second line compiles the AC(0.8) results complemented by imposing the $\alpha_{\mu\nu}(z^*)=\alpha_{\nu\mu}(z)$ symmetry that forces the poles of the continued-fraction to be real. We provide 3 decimal places for sake of comparison. } 
\label{tab:tableC60}
\end{table}

By construction, the ${\Im} \alpha(z)$ curve for $(z  =\omega +i\Gamma)$ goes exactly through the ${\Im} \alpha(z_k)$ data points (black diamonds). Concerning the continuation to the $(\omega + i\gamma)$ energy line ($\gamma$=0.2~eV), the agreement with the reference (pink) function is  excellent, except for the highest peak at $\sim$6.8~eV. The analysis of the poles of the continued-fraction representation indicates that the corresponding peaks land at 3.813~eV, 4.948~eV, 6.038~eV and 6.890~eV, namely an error between 20 and 40 meV for the 3 lowest main peaks, and a maximum error of 0.11~eV for the highest  energy peak [see AC(0.8) line in Table~\ref{tab:tableC60}]. Further, the poles and the associated residues are associated with a small imaginary part. For instance, the last pole appears really at ($6.890+i0.107$) eV. We speculate that this latter feature lies close to the strongest peak at $\sim$6.08~eV, inducing potentially difficulties to resolve it. The poles of the continued-fraction are 3-fold degenerate, indicating faithfully the degeneracy of the  bright excitations. 

\begin{figure}[t]
\begin{center}
 	\includegraphics[width=0.49\textwidth]{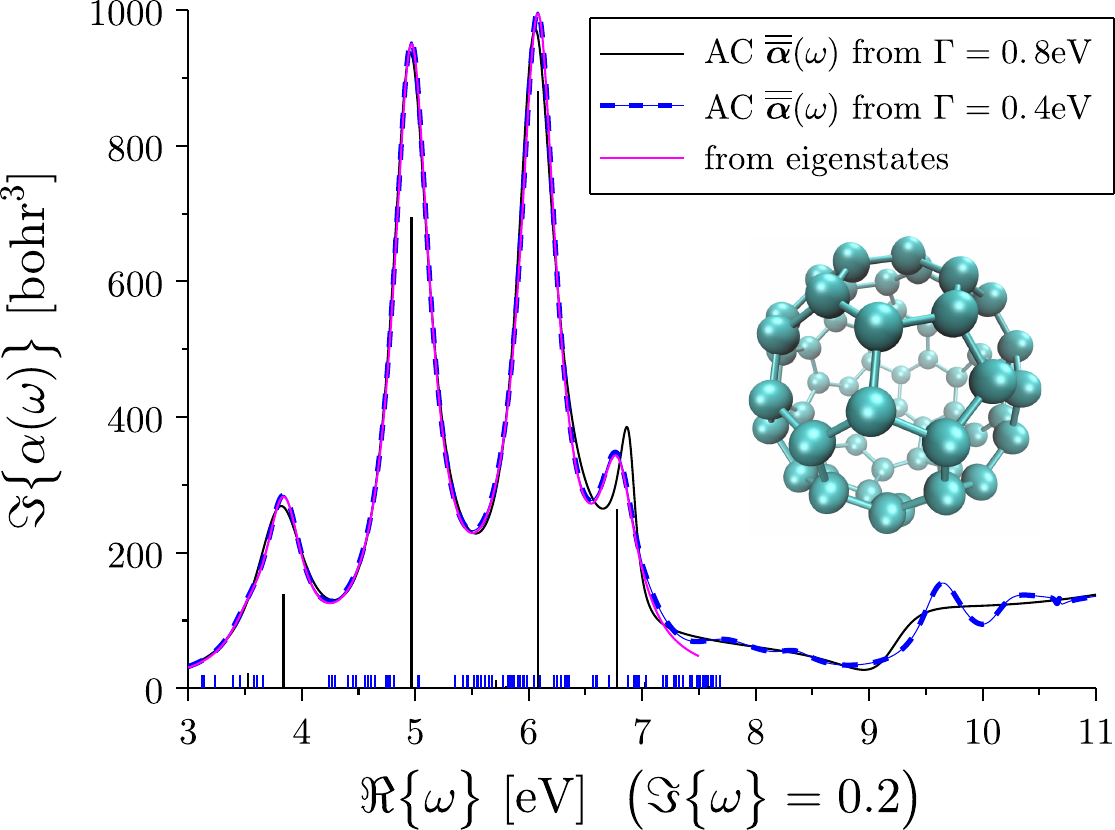}
    \end{center}
	\caption{ Same as Fig.~\ref{fig:figc60} adding in blue dashed the results obtained with a finer   sampling grid in the complex plane [AC(0.4) scheme].}
	\label{fig:figc60bis}
\end{figure}

The agreement with the reference data can be significantly improved by adopting a tighter grid of reference ($z_k$) frequencies in the complex plane using an improved ``AC(0.4)" scheme. Namely, we adopt a grid $(z_k =  \omega_0 + k\Delta\omega +i\Gamma)$ with $\Gamma=1.5\Delta\omega=0.4$~eV, that is twice as many sampling frequencies in the complex plane, getting closer to the $(\omega+i\gamma$)  energy line. 
The resulting absorption spectrum is represented with the dashed blue line in Fig.~\ref{fig:figc60bis}. The agreement with the reference is now excellent, as indicated further by the energies and oscillator strengths reported in Table~\ref{tab:tableC60} [AC(0.4)  line] as extracted from the poles and residues of the corresponding continued-fraction representation.  

While the use of a finer sampling grid in the complex plane results as expected in better results, we provide a strategy for improving the results associated with the very coarse AC(0.8) sampling grid. In the case of the $C_{60}$  fullerene, the continued-fraction representation attempts to reproduced isolated peaks with a pole energy on the real-axis. In such a situation, one can enforce the 
 $\bar{\bar{\alpha}}(z_k^*) = \bar{\bar{\alpha}}(z_k)^{*}$   symmetry. This doubles at no cost the number of reference frequencies, resulting in a continued-fraction representation with twice as many coefficients, forcing further the poles to lie on the real axis. 
 The analysis of the poles of this updated tensorial continued-fraction representation allows locating exactly the 4 lowest peaks energy at 3.839~eV (f=0.162), 4.966~eV (f=0.813), 6.080~eV (f=1.028) and 6.781~eV (f=0.313), in   agreement at the meV level with the BSE eigenstates reference (see the  "+ $\alpha(z^*)$" line in Table~\ref{tab:tableC60}). The oscillator strengths also agree very nicely with the reference.   Such a strategy only slightly improves the results associated with the finer  AC(0.4)  sampling grid  yielding results in already  excellent agreement with the reference data.

\subsection{ The PCBM derivative }

\begin{figure}[t]
\begin{center}
 	\includegraphics[width=0.49\textwidth]{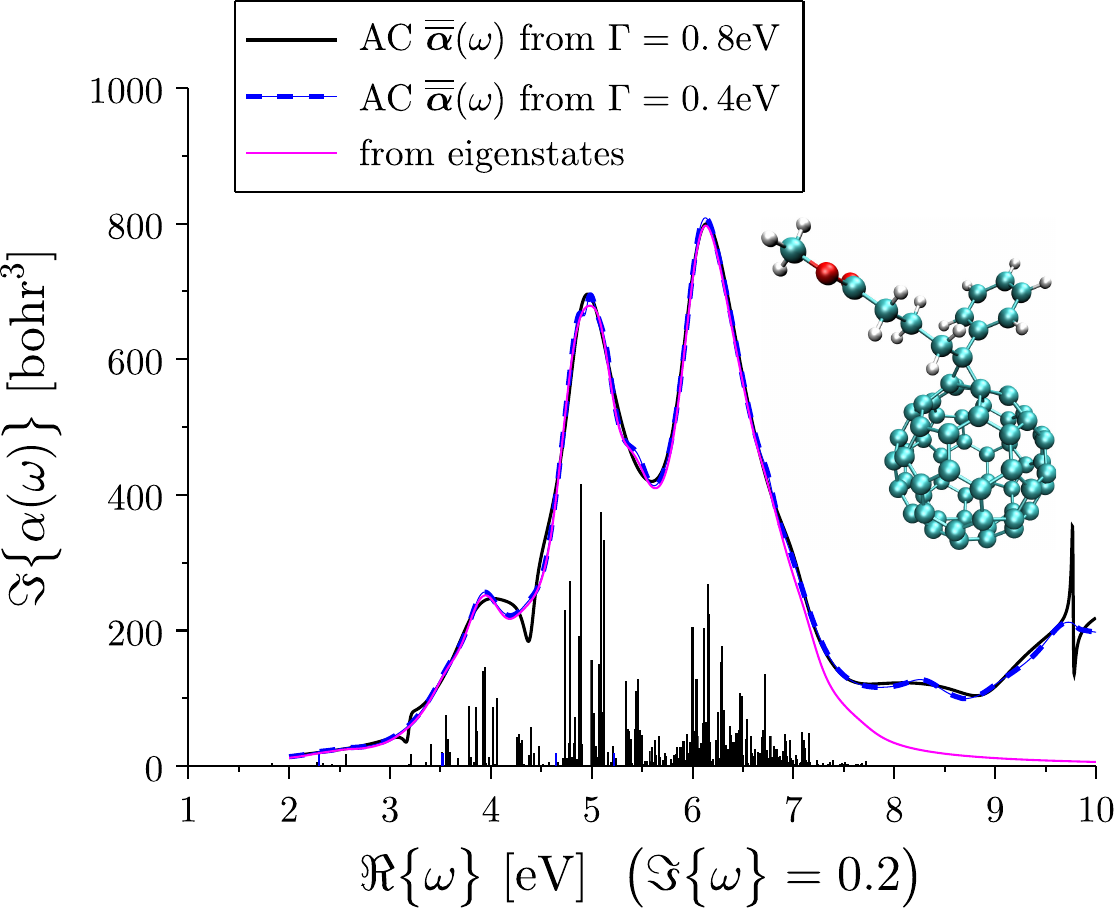}
    \end{center}
	\caption{ Same as Fig.~\ref{fig:figc60bis} but for the PCBM derivative.}
	\label{fig:figPCBM}
\end{figure}

We now address the more complex case of the [6,6]-phenyl-C61-butyric acid methyl ester (PCBM) derivative (see molecular representation Inset Fig.~\ref{fig:figPCBM})  that stands as a very popular acceptor in organic solar cells, even though challenged by a new generation of non-fullerene acceptors. Due to the presence of the grafted side-chain, all symmetries are broken and the dark excitations are now brightened, resulting in a more complex absorption spectrum. This is illustrated in Fig.~\ref{fig:figPCBM}  where we plot the 400 lowest BSE eigenstates (vertical bars) and in pink the reference  averaged trace $\Im \alpha(\omega+i\gamma)$ reconstructed from the available BSE eigenstates with  $\gamma$=0.2~eV. \red{ Here again we perform BSE/ev$GW$@PBE0 cc-pVTZ calculations on the B3LYP cc-pVTZ geometry (provided in the SM). }

The PCBM spectrum resembles that of C$_{60}$ but with broadened features, with the highest peak at 6.8~eV merging as a weak shoulder of the main peak at about 6.2 eV.  Contrary to the fullerene case with well-resolved individual transitions, the small  ``{e}xperimental"  broadening ($\gamma$=0.2~eV) merges closely lying transitions, resulting in broad structures that serve as an envelope to a large number of BSE eigenstates. As discussed above, attempting to resolve individual eigenstates with an analytic continuation scheme is probably hopeless. Further, poles of the continued-fraction representation account for several closely lying individual excitations, so that the associated    energy and residue cannot be associated, as in the case of the pristine $C_{60}$ fullerene, with individual transitions.

Considering the blue-dashed line in Fig.~\ref{fig:figPCBM}, we observe that despite the complexity of the spectrum, the analytic continuation built with the AC(0.4) scheme  allows capturing faithfully the reference spectrum.   This represents calculating 32   BSE $\bar{\bar \alpha}(z_k)$ tensors with $\Re z_k$ within [2,10]~eV.    As such, selecting a grid with $\Gamma=2 \gamma$ and $\Delta\omega=\Gamma/1.5 \simeq0.27$~eV seems again to stand as a safe choice. 

\begin{figure}[t]
\begin{center}
 	\includegraphics[width=0.49\textwidth]{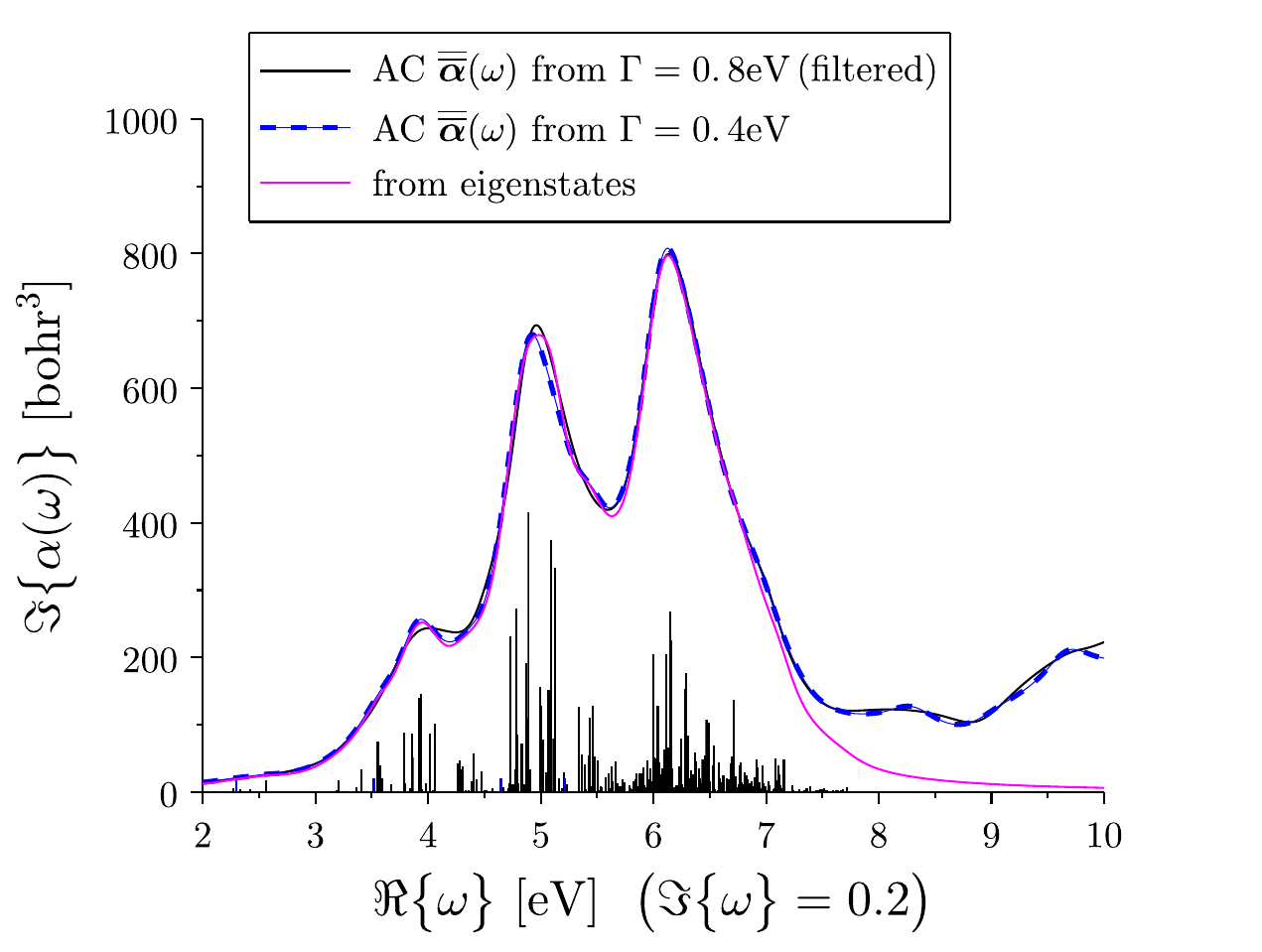}
    \end{center}
	\caption{ Same as Fig.~\ref{fig:figPCBM} but setting to zero the imaginary part of the 3 poles that have entered the first-quadrant, yielding the AC(0.8)-filtered scheme.  }
	\label{fig:PCBM2}
\end{figure}

Considering now the spectrum associated with the rough AC(0.8) scheme, calculating twice as less reference $\bar{\bar \alpha}(z_k)$ tensors, we observe an overall good agreement, but with small instabilities appearing at  3.2 eV,  4.4 eV and 9.8 eV. Analysis of the continued-fraction reveals indeed poles with a real-part at these energies. More importantly, they are located in the first quadrant (positive imaginary part)  close to the ($\omega+i\gamma$) energy line.  Even though  associated with very small residues, these poles create locally a weak to significant deviation from the reference spectrum. The other poles are located in the second-quadrant as expected to reproduce a positive $\Im \alpha(\omega)$. We speculate that the pole at 4.4~eV originates from attempting to reproduce the small dip at $\sim$4.2~eV present in the true spectrum. Setting the imaginary-part of these poles to zero leads to a regularized rational form (Eq.~\ref{eqn:eqpoles}) with reduced singularities (see Fig.~\ref{fig:PCBM2}, ``{f}iltered" results).

\subsection{ Ag$_{20}$ surface plasmon resonances}

The optical properties of noble metal clusters are characterized by the emergence with size of a strong absorption line in the near-UV/visible range. These modes can be associated with localized surface plasmon  resonances (LSPR) yielding very large induced fields in their vicinity, leading to numerous applications. \cite{Schuller2010} Concerning silver clusters, such modes have been shown to gradually develop with increasing cluster size into a broad unique structure in the [3.9-4.1] eV range, originating from the merging of a very large number of excitations. \cite{Yu2018} Recently,  we found that the BSE formalism can predict accurately the absorption spectra of small Ag$_n$ clusters (n=2,4,6,8) provided that (a) the 4\textit{s} and 4\textit{p} shells are not frozen in the pseudopotential (or Empirical Core Potential) and (b) that self-consistency at the $GW$ level is accounted for when starting with DFT eigenstates generated with (semi)local functionals. \cite{Blase2025}

\begin{figure}[h]
\begin{center}
 	\includegraphics[width=0.45\textwidth]{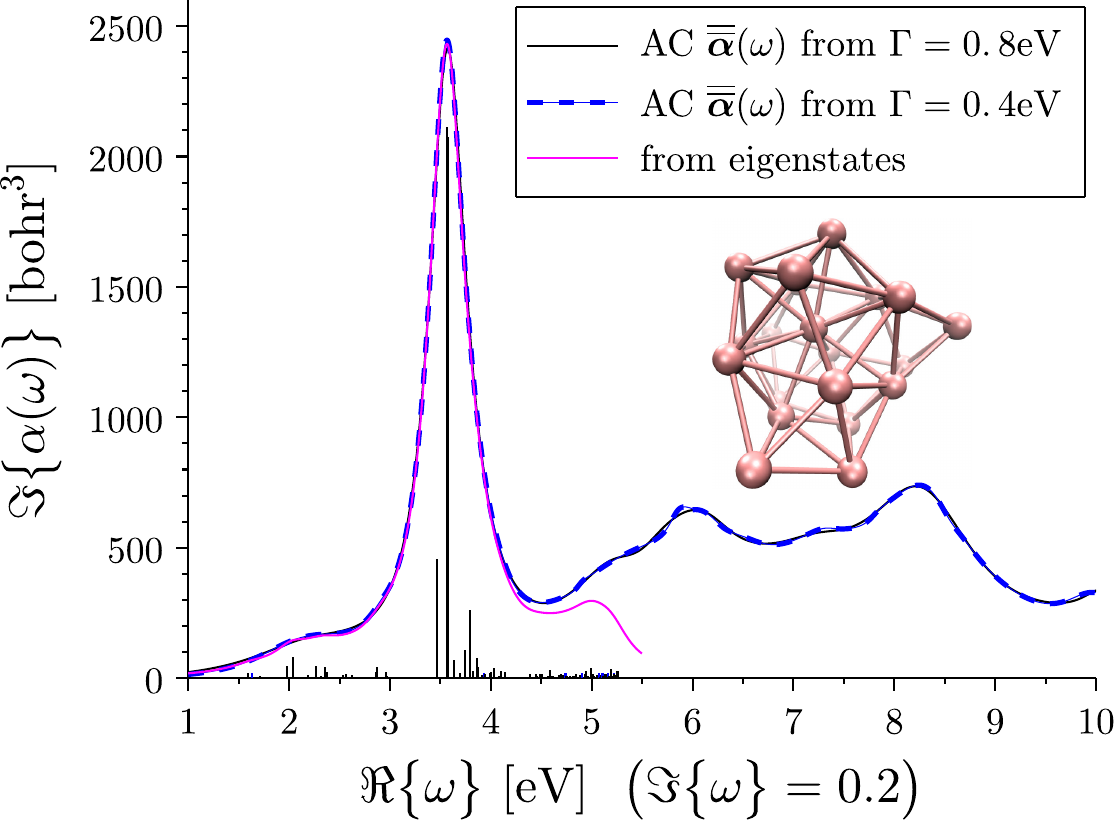}
    \end{center}
	\caption{ Same as Fig.~\ref{fig:figc60bis}  but for the Ag$_{20}$ metallic cluster. The vertical bars are associated with the 280 lowest BSE eigenstates energy and oscillator strength.
}
	\label{fig:figAg20}
\end{figure}

We plot in Fig.~\ref{fig:figAg20} the BSE/ev$GW$@PBE def2-TZVP $\; \Im \alpha(\omega) = \Im[  Tr \; \bar{\bar \alpha}(\omega)]/3 \;$ spectrum associated with the Ag$_{20}$ cluster, adopting the lowest energy isomer as found in Ref.~\citenum{Nhat2018}. The Ag$_{20}$ cluster was shown to be of the size for which an LSPR mode  starts emerging. \cite{Yu2018} We calculate explicitly 280 BSE eigenstates.
The BSE absorption spectrum is dominated by 3 nearly-degenerate excited-states at $\sim$3.57 eV ($f{\sim}1.3$) that are the 60$^{th}$, 61$^{st}$  and 62$^{nd}$ BSE eigenstates. With increasing size, the number of states contributing to the resonance increases into a very dense ``forest" of states, merging into a quasi-continuum with an envelope that defines the resonance energy distribution. 

The BSE eigenstates are used to reconstruct   $\;\Im  \alpha(\omega+i\gamma) \;$  with $\gamma$=0.2~eV, standing  as the reference  (pink line). This is again compared to the result of using the rough AC(0.8) scheme (solid black line) and the refined AC(0.4) approach (dashed blue line). The agreement with the reference absorption curve is excellent in both cases, except above 4.5~eV where the lack of available BSE eigenstates above $\sim$5.2~eV depletes the reconstructed (pink) spectrum. Further, in a large energy range above the highest available reference BSE eigenstate, the two AC functional forms yield  very similar results, indicating the convergence of the two sampling grids. As such, the present scheme stands as a very valuable tool to explore efficiently the surface plasmon resonance of larger clusters and the related exaltation of the absorption lines of nearby molecules.

\subsection{ Core levels }

The case of core levels is typical of situations where the targeted excitations lie far away in energy from the lowest valence excitations. In such situations,  standard eigenvalue solvers starting from the lowest (valence) excitation energies may reveal impracticable in the case of increasingly large systems. To bypass this difficulty, a core-valence-separation (CVS) approximation may be adopted, removing  the valence eigenstates from the (occupied)$\times$(virtual) product space over which the TD-DFT or BSE Hamiltonian is constructed. This turns to be a good approximation for the absorption by deep core levels located hundreds of eV below the valence occupied states. \cite{Urquiza2023,Kehry2023} However, even with the CVS approximation, the number of transitions from core levels in a given energy window grows quadratically with the size of the system. 

As compared to the valence excitations discussed here above, the case of core levels presents no additional difficulties when it comes to use the present AC scheme. If targeting excitations in some [$\omega_{min} ,\omega_{max}$]  energy range, the only modification is to take the $ (  z_k  = \omega_0 + k\Delta\omega + i\Gamma)$ sampling grid such that the real-part of the ($z_k$) complex frequencies covers this targeted energy-range. The constructed continuous-fraction functional form will be used to reconstruct the absorption spectrum in this energy range. We emphasize further that even in the case of the valence excitations treated above, the BSE product-space was built using all core, valence     and virtual states. As such, the GMRES scheme, with successive Hamiltonian-on-vector projections in this product-space, presents the very same cost whatever the targeted energy range. We further did not observe differences in the number of needed iterations. 

As a proof-of-concept illustrative case, we consider again the $C_{60}$ fullerene for which experimental   X-ray absorption spectra (XAS) are available. \cite{Luo1995}  We  adopt the same \red{cc-pVTZ} basis set used for valence excitations. We do not attempt  to uncontract the contracted channels as suggested for core levels ionization energy $GW$ calculations. \cite{Mejia2022}  Likewise, we do not include relativistic corrections. \cite{Keller2020,Golze2020,Yao2022,Kehry2023} Our goal here is to compare the absorption spectrum reconstructed from   explicit  eigenstates  BSE calculations with the present AC scheme, all calculations using the very same running parameters. Our preceding $GW$ calculations exploit the improved analytic continuation scheme developed for core levels. \cite{Duchemin2020,Barrueta2023,Kehry2023}

\begin{figure}[t]
\begin{center}
 	\includegraphics[width=0.45\textwidth]{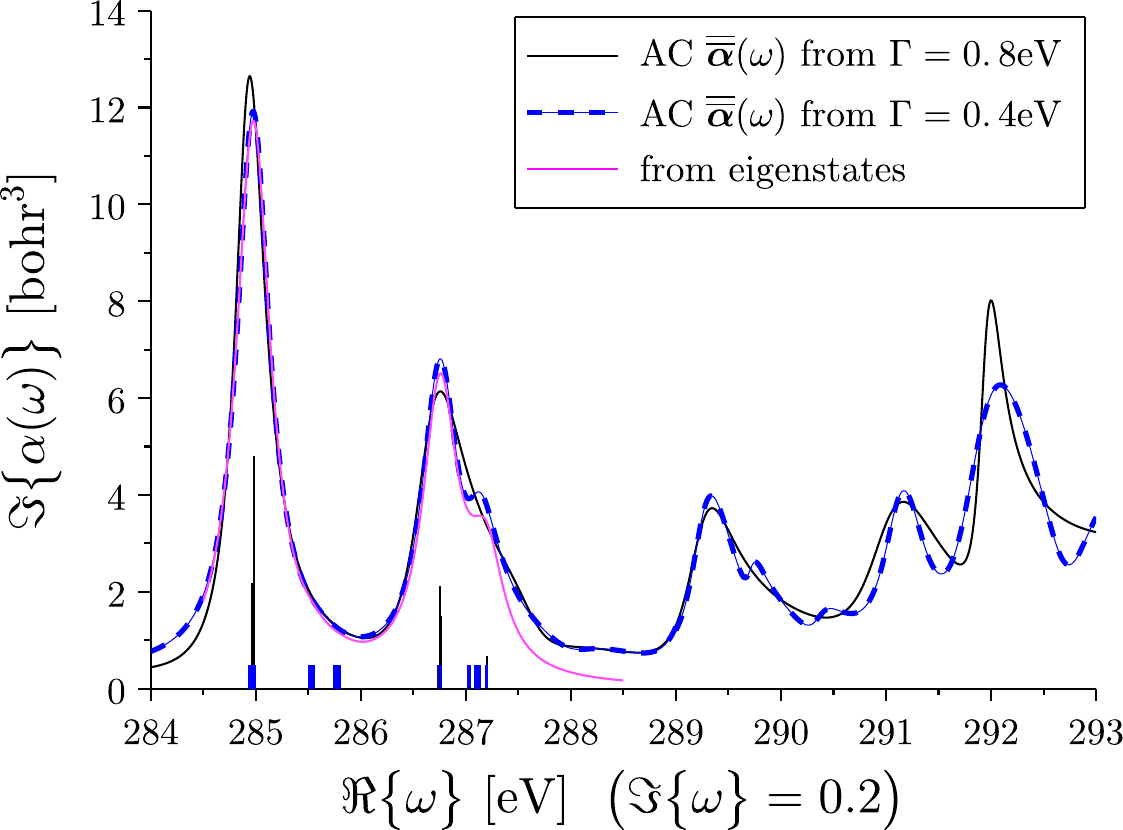}
    \end{center}
	\caption{ Same as Fig.~\ref{fig:figc60bis}  but for the C$_{60}$ fullerene C$_{1s}$ core levels excitations. The vertical bars correspond to the  400  lowest BSE excitation energies calculated with the CVS approximation.  }
	\label{fig:figcore}
\end{figure}

We plot in Fig.~\ref{fig:figcore} the C$_{60}$ fullerene  $\Im \alpha(\omega+i\gamma)$ average spectrum ($\gamma$=0.2~eV) in the [284,293]~eV energy range. The pink line represents the reference spectrum as reconstructed by the explicitly calculated  BSE eigenstates (vertical black bars). Due to computational cost, the CVS approximation was implemented and only  400  eigenstates were calculated. The lowest peak  at $\sim$284.96~eV  is within 0.6~eV of the lowest 284.4~eV   peak in the experimental XAS spectrum. \cite{Luo1995}  The double structure at 286.7~eV/287.2~eV is within $\sim$0.9~eV of   the 285.8~eV/286.3~eV  experimental double-peak.  As stated above, the lack of relativistic corrections and of a proper basis set adapted to core levels, may easily explain  such a discrepancy, together with the BSE/ev$GW$@PBE0 scheme in itself (for recent benchmarks of the BSE scheme on molecular XAS spectra, see Ref.~\citenum{Yao2022}).

We now consider the analytic continuation scheme. We  calculate the reference $\bar{\bar \alpha}(z_k)$ tensors with and without the CVS approximation. Differences are trifling, indicating that for the fullerene XAS spectrum, the CVS approximation is very robust, confirming recent studies at the BSE level. \cite{Urquiza2023,Kehry2023} The continued-fraction functional form built with the AC(0.4) scheme (dashed blue line) yields again a  good agreement with the reference (pink) spectrum. The two first peaks are within 10 meV of the corresponding BSE eigenstates. Deviations start to occur a few tenths of an eV before the highest explicitly calculated BSE excitation due to the lack of contributions from missing higher lying energy excitations in the pink-line reference. 
One verifies that using the $\bar{\bar{\alpha}}(z^*) = \bar{\bar{\alpha}}(z)^*$ symmetry hardly changes the position of the peaks over the entire energy range, indicating good convergence for the AC(0.4) scheme.

The results associated with the analytic continuation built from the rough AC(0.8) scheme (black full line)  are also in good agreement with the reference (pink) data, even though missing the weak shoulder at 287.2 eV. Similarly, comparing with the  robust AC(0.4) approach, the continued-fraction form build with the very-coarse grid provides a good description of the main spectrum features above the energy range where explicit BSE eigenstates are available, but misses the small structures at  289.75~eV and 290.5~eV. 

The  results presented in this Section illustrate that  core excitations can be treated without any difference in terms of cost and accuracy as compared to valence excitations, without any need to implement the CVS approximation. In particular, the sampling grid associated with the AC(0.4) approach offers again a good compromise between efficiency and accuracy. Such a statement relies on the efficient  and similar analytic continuation approach to calculating upstream the needed $GW$ quasiparticle energies in the core region, as implemented and tested previously by several groups.~\cite{Duchemin2020,Barrueta2023,Kehry2023}

\section{Conclusion}

We have explored the possibility to build the BSE absorption spectrum in selected energy windows on the basis of calculating iteratively, with $\mathcal{O}(N^{4})$ scaling, the BSE polarizability (3$\times$3) tensors $\bar{\bar \alpha}(z_k)$ for a coarse grid of   $\lbrace z_k \rbrace$ frequencies in the complex plane. These data can be used to build an $\bar{\bar \alpha}^{AC}(z)$ tensorial continued-fraction representation of the polarizability tensor. The resulting analytic functional form allows calculating $\bar{\bar \alpha}^{AC}(\omega+i\gamma)$   close to the real-energy axis (small $\gamma$ broadening) in an energy range corresponding roughly to the extent of the real part of the selected $\lbrace z_k \rbrace$ complex frequencies. In the examples we explored, the number of 
$\bar{\bar \alpha}(z_k)$ tensors to be calculated amounts to about 20 to 40 ($z_k)$ frequencies to span a 10~eV energy range along the real-axis. Both valence and core excitations were explored in the case of the C$_{60}$ fullerene. 

The limitation of the present approach is that building a continued-fraction representation of finite order can only capture a finite number of poles.  Calculating $N_f$ $\bar{\bar \alpha}(z_k)$ reference $3\times3$ tensors allows the capture of a maximum of $3\times N_f/2$ poles in the vicinity of the energy range spanned by the real-part of the ($z_k$). 
As such, the present approach will tend to merge closely lying peaks if the number of bright excitations in the chosen energy range exceeds the continued fraction capacity. This may be desirable in the case of, e.g., plasmon resonances composed of a very large number of closely lying excitations, but may stand as a limitation whenever closely lying excitations need to be resolved. 
If needed, accumulating a few more sampling frequencies in this energy range may help resolving more structures, suggesting an adaptative grid strategy beyond the simple uniform grid explored here above.

\section*{SUPPLEMENTARY MATERIAL}
See the Supplementary Material for   an illustration of the difficulties associated with performing a scalar analytic continuation for selected components of the $\alpha_{\mu\nu}$ tensor  in the case of the $\beta$-dipeptide, \red{together with the  $C_{60}$ and  PCBM derivative   B3LYP cc-pVTZ geometries used in the present study. Further, all data to reconstruct the continued-fraction for the $\beta$-dipeptide in the AC(0.8) scheme are provided. }  

\begin{acknowledgments}
XB acknowledges Francesco Sottile for discussions on core levels X-ray absorption spectroscopy BSE calculations in the case of inorganic periodic systems.
This project was provided with HPC computing and storage resources by GENCI at CINES and TGCC  thanks to the grants 2025-A0110910016 and 2025-AD010916992 on the supercomputers  Joliot Curie's  ROME and Adastra's GENOA  partitions, respectively.
\end{acknowledgments}

\section*{Data Availability Statement}

Most data that support the findings of this study are available within the article  and its supplementary material. Additional data, such as the polarizability tensors and matrix-valued $\bf b$-coefficients for other systems than the $\beta$-dipeptide, available in the Supplementary Material, can be provided on reasonable demand. 

\vskip 2cm

%


\end{document}